\documentclass{article}
\usepackage{ijcai24}

\usepackage{times}
\usepackage{soul}
\usepackage{url}
\usepackage[hidelinks]{hyperref}
\usepackage[utf8]{inputenc}
\usepackage[small]{caption}
\usepackage{graphicx}
\usepackage{amsmath}
\usepackage{amsthm}
\usepackage{booktabs}
\usepackage{algorithm}
\usepackage{algorithmic}
\usepackage[switch]{lineno}
\usepackage[cjk]{kotex}
\usepackage{amsfonts}
\usepackage{multirow}
\usepackage{mathtools}
\usepackage{graphicx, wrapfig}
\usepackage{subfigure}
\usepackage{optidef}
\usepackage{bm}
\usepackage{float}
\usepackage{siunitx}
\usepackage{lipsum}
\usepackage{pifont}
\newcommand{\cmark}{\ding{51}}%
\newcommand{\xmark}{\ding{55}}%
\usepackage[normalem]{ulem}
\useunder{\uline}{\ul}{}
\usepackage[textsize=footnotesize]{todonotes}
\usepackage{marginnote}
 \let\marginpar\marginnote
\newtheorem{theorem}{Theorem}

\title{SVD-AE: Simple Autoencoders for Collaborative Filtering}

\author{
Seoyoung Hong$^1$\thanks{Work done while at Yonsei University.}\and
Jeongwhan Choi$^2$\and
Yeon-Chang Lee$^3$\and
Srijan Kumar$^4$\And
Noseong Park$^5$\thanks{Corresponding author.}\\
\affiliations
$^1$Boeing Korea Engineering and Technology Center (BKETC)\\
$^2$Yonsei University\\
$^3$Ulsan National Institute of Science and Technology (UNIST)\\
$^4$Georgia Institute of Technology\\
$^5$Korea Advanced Institute of Science and Technology (KAIST)\\
\emails
sydney.hong@boeing.com,
jeongwhan.choi@yonsei.ac.kr,
yeonchang@unist.ac.kr,
srijan@gatech.edu,
noseong@kaist.ac.kr
}

\begin{document}

\maketitle

\begin{abstract}
Collaborative filtering (CF) methods for recommendation systems have been extensively researched, ranging from matrix factorization and autoencoder-based to graph filtering-based methods. Recently, lightweight methods that require almost no training have been recently proposed to reduce overall computation. However, existing methods still have room to improve the trade-offs among accuracy, efficiency, and robustness. In particular, there are no well-designed closed-form studies for \emph{balanced} CF in terms of the aforementioned trade-offs. In this paper, we design SVD-AE, a simple yet effective singular vector decomposition (SVD)-based linear autoencoder, whose closed-form solution can be defined based on SVD for CF. SVD-AE does not require iterative training processes as its closed-form solution can be calculated at once. Furthermore, given the noisy nature of the rating matrix, we explore the robustness against such noisy interactions of existing CF methods and our SVD-AE. As a result, we demonstrate that our simple design choice based on truncated SVD can be used to strengthen the noise robustness of the recommendation while improving efficiency. Code is available at \url{https://github.com/seoyoungh/svd-ae}.
\end{abstract}

\section{Introduction}

Recommender systems have been utilized extensively in a variety of service platforms, including e-commerce and social media~\cite{ying2018graph,sharma2022survey,shin2024bsarec}. Collaborative Filtering (CF), a crucial task for recommendation, typically learns latent vectors (i.e., embeddings) of users and items from their prior user-item interactions and provides personalized preferred recommendations~\cite{rendle2012bpr,he2017neural}. Since the user-item interaction matrix can be represented as a binary graph, graph-based CF, which learns user and item embeddings using graph convolutional neural networks (GCNs), has been widely studied recently~\cite{he2020lightgcn,lee2021bootstrapping,choi2021lt,kong2022hmlet,hong2022timekit,choi2023rdgcl,kim2023trustsgcn,lee2024stochastic}.

\begin{table}[t]
    \small
    \setlength{\tabcolsep}{2pt}
    \centering
    \begin{tabular}{@{}lcccc@{}}
        \toprule
        & GF-CF & EASE & $\infty$-AE & \textbf{SVD-AE} \\ \midrule
        Closed-form Solution        & \cmark & \cmark & \cmark & \cmark \\
        Autoencoder-based           & \xmark & \cmark & \cmark & \cmark \\
        Using SVD                   & \cmark & \xmark & \xmark & \cmark \\
        Using Neural Networks       & \xmark & \xmark & \cmark & \xmark \\
        Inference Target & Item-item & Item-item & User-user   & User-item  \\
        \bottomrule
    \end{tabular}
    \caption{Comparison of existing lightweight methods~\protect\cite{shen2021powerful,steck2019embarrassingly,sachdeva2022infinite} and our SVD-AE.}
    \label{tbl:cmp1}
\end{table}

At the same time, the escalating expense of training models on massive datasets, including billions of user-item interactions, has become a difficulty in the study of recommender systems. 
Since LightGCN~\cite{he2020lightgcn} simplified the GCN-based CF and improved its performance by eliminating nonlinear activation functions and feature transformations, recent works have attempted to make LightGCN more lightweight~\cite{mao2021ultragcn,peng2022svd}.
Despite these advancements, learning-based approaches still have drawbacks in that they need a lot of time and computation since they require training. 

Therefore, various research has provided lightweight and effective computing approaches based on closed-form solutions to lessen the computational complexity~\cite{steck2019embarrassingly,shen2021powerful,sachdeva2022infinite,choi2023bspm}. GF-CF~\cite{shen2021powerful} designs a simple and computationally efficient framework that leverages various graph filters. On the other side, autoencoders (AE) are also used for constructing the closed-form solution for CF. EASE~\cite{steck2019embarrassingly} discovers the closed-form solution of the shallow AE-based recommender system by solving its linear regression-based formulation via the method of Lagrangian multipliers. $\infty$-AE~\cite{sachdeva2022infinite} suggests adopting the neural tangent kernel (NTK) to obtain the closed-form solution of the autoencoder with an infinitely-wide bottleneck layer for recommendation. 

While these methods exhibit successful considerations for efficiency and performance in their design, they face challenges in achieving excellent generalization abilities due to the prevalent presence of noisy interactions in the recommendation data.
It is well-known that a user-item interaction matrix, $\mathbf{R} \in \{0,1\}^{|U| \times |I|}$, where $U$ is a set of users and $I$ is a set of items, often contains noise~\cite{said2012users}.
That is, an interaction
does not necessarily mean that a user is satisfied with an item~\cite{amatriain2009rate,sar2015data}. 
In this case, EASE and $\infty$-AE employ straightforward reconstruction of 
$\mathbf{R}$,
which can lead to noisy interactions if not properly controlled~\cite{steck2020autoencoders}.


In this work, we aim to explore the way for the \emph{best overall balance between accuracy, computational speed, and noise robustness}.
Toward this goal, we propose an \textbf{SVD-AE} for CF, whose optimal solution can be analytically defined with the truncated SVD, to effectively exclude the noise in the original rating matrix when inferring $\hat{\mathbf{R}}$. 
Additionally, our SVD-AE directly reveals $\hat{\mathbf{R}}$ from $\mathbf{R}$ without any intermediate steps, thus reducing the computation time, whereas other methods, e.g., GF-CF, EASE, and $\infty$-AE, require obtaining the item-item or user-user similarity matrix from $\mathbf{R}$. 
We experimentally show that our simple design choice based on truncated SVD leads to a practical way to i) enhance the balance among the recommendation accuracy, the computation time, and the robustness, and ii) enable obtaining a closed-form solution devoid of any parameters to train. 
In \textbf{Table~\ref{tbl:cmp1}}, we compare our SVD-AE with existing lightweight methods, which will be discussed in more detail in Sections~\ref{sec:discussion} and~\ref{sec:relation}.


\begin{figure}[t]
    \centering
    \includegraphics[width=0.68\columnwidth]{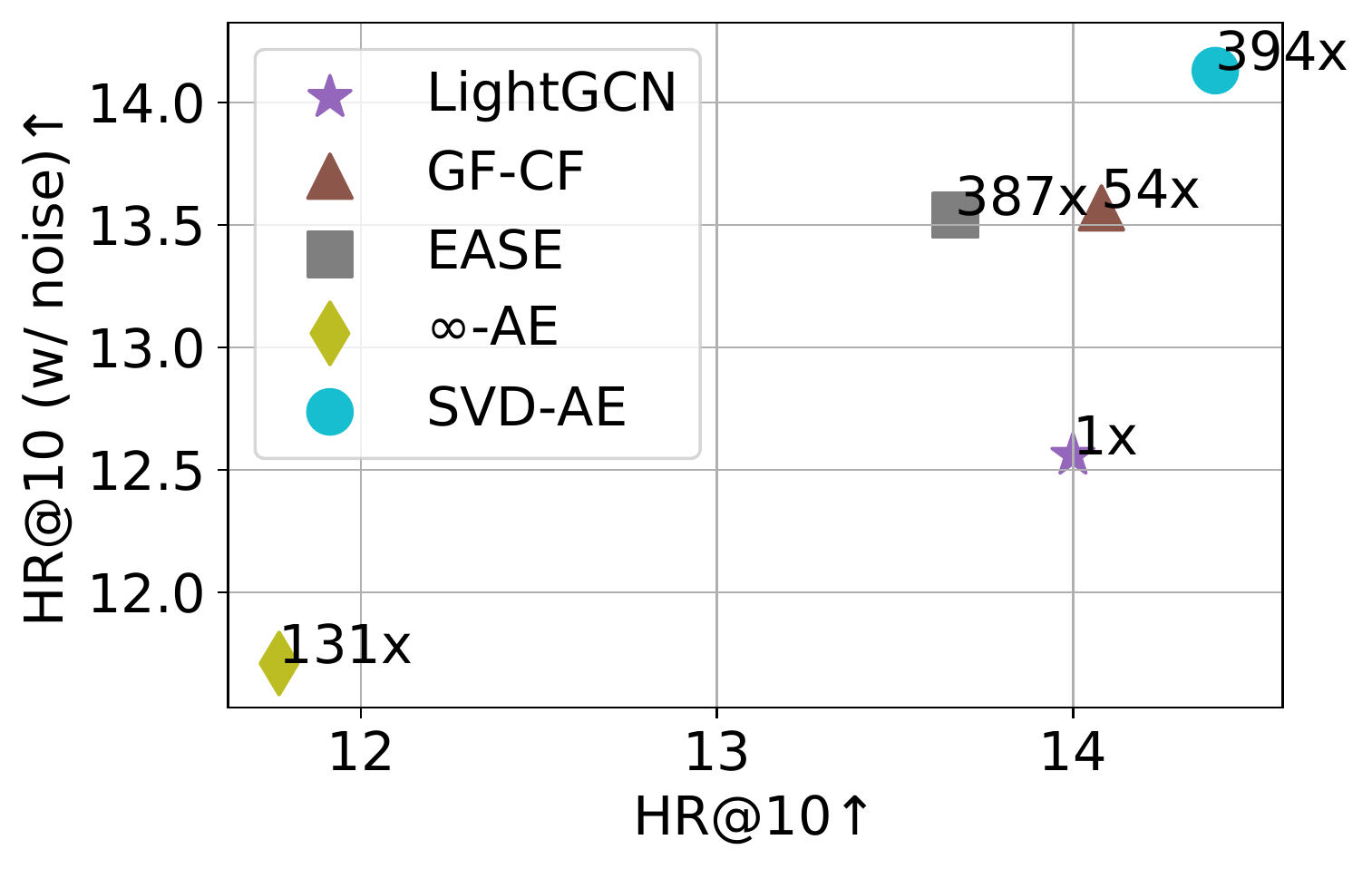} 
    \caption{The accuracy, robustness, and computation time of various methods on Gowalla. The \boldsymbol{$x$}-axis indicates each method's accuracy on the original dataset, and the \boldsymbol{$y$}-axis indicates each method's robustness, i.e., its accuracy on the dataset with 5\% random interactions added out of all user-item interactions.
    The computation time is indicated by how fast each method is compared to LightGCN (1x).}
    \label{fig:overview_vis}
\end{figure}

The experiments with 8 baselines on 4 datasets demonstrate that our SVD-AE marks the best place for HR@10, NDCG@10, and PSP@10 by non-trivial margins. 
Furthermore, we study the robustness against the addition of random noisy interactions of all the baselines and our SVD-AE. 
As shown in \textbf{Figure~\ref{fig:overview_vis}}, SVD-AE has the fastest computation time (spec., 394 times quicker than LightGCN), provides the best recommendation accuracy (HR@10), and performs best even with the addition of random noisy interactions, demonstrating its robustness. 

Our contributions can be summarized as follows: 
\begin{enumerate}
    \item \textbf{Novel Closed-form Solution for CF:} Our SVD-AE, a simple but efficient linear AE using SVD, has a closed-form solution for CF. We also devise a generalized AE form that subsumes all existing AE-based CF methods.
    \item \textbf{Robust Methodology:} 
    We discuss how to use the truncated SVD to avoid potential overfitting by the noisy interactions. Through in-depth analyses, we also investigate the superior robustness of AE-based recommendation methods to noise, including our SVD-AE.
    \item \textbf{Comprehensive Validation:} Our method outperforms all baselines in all benchmark datasets for HR@10, NDCG@10, and PSP@10. Furthermore, compared to many other existing methods, the computation time is much shorter as it has an even simpler closed-form solution than others. Therefore, our method offers the best overall balance between the accuracy of the recommendation, the computation time, and the noise robustness.
\end{enumerate}

\section{Related Work \& Preliminaries}

\subsection{Lightweight Collaborative Filtering}
In recent years, several CF methods have been developed to build lightweight recommender systems~\cite{shen2021powerful,choi2023bspm,park2024turbo}.
First, GF-CF~\cite{shen2021powerful} designed a simple and computationally efficient framework that leverages various graph filters. 
Other methods typically employ autoencoder approaches based on linear regression~\cite{steck2019embarrassingly,sachdeva2022infinite}. EASE~\cite{steck2019embarrassingly} suggested a shallow autoencoder that simply calculates an item-item similarity matrix using ordinary least squares regression with a closed-form solution. They also used Lagrangian multipliers to constrain the self-similarity of each item in the input and output layer to zero. Recently, $\infty$-AE~\cite{sachdeva2022infinite} adopted the NTK to obtain the closed-form solution of the autoencoder with an infinitely-wide bottleneck layer for recommendation. RLAE and RDLAE~\cite{moon2023s} extended EASE by adjusting the degree of diagonal constraints through tuning for L2 regularization with random dropout.



\subsection{Generalized Linear Autoencoder for Recommender Systems}
In this section, we describe a generalized linear autoencoder for recommender systems and demonstrate how simple regression methods solve the problem of finding the optimal model. We then present existing methods to solving the linear regression.

The objective function of linear autoencoder is written as:
\begin{align}\label{eq:generalized_ae}
    \operatorname*{min}_{\hat{\mathbf{R}}}&\quad \| \mathbf{R} - \hat{\mathbf{R}} \|_2^2, \quad
    \textrm{s.t.} \quad \mathcal{C},
\end{align} where $\mathcal{C}$ denotes a set of constraints. $\mathbf{R} \in \{0,1\}^{|U| \times |I|}$ and $\hat{\mathbf{R}} \in \{0,1\}^{|U| \times |I|}$ represent a given user-item interaction matrix and a reconstructed interaction matrix, respectively.
$U$ and $I$ denote sets of users and items, respectively.

Many existing recommendation methods can be generalized into the above form with specific choices for their own constraints. Their closed-form solutions for optimal $\hat{\mathbf{R}}$ are given by:
\begin{equation}\label{eq:r_hat}
  \hat{\mathbf{R}} =
    \begin{cases}
      \mathbf{R} \cdot (\mathbf{I} - \hat{\mathbf{P}} \cdot \text{diagMat}(\vec{1} \oslash \text{diag}(\hat{\mathbf{P}}))) & \text{(EASE)}, \\
      \mathbf{K} \cdot (\mathbf{K} + \lambda \mathbf{I})^{-1} \cdot \mathbf{R} & \text{($\infty$-AE)}, \\
      \tilde{\mathbf{R}} \cdot \mathbf{V} \tilde{\mathbf{\Sigma}}^{+}\mathbf{Q}^T\mathbf{R} & \text{(SVD-AE)},
    \end{cases}       
\end{equation} where the specific interpretations are:
\begin{enumerate}
    \item In the case of EASE~\cite{steck2019embarrassingly}, $\mathbf{\hat{P}} = (\mathbf{R}^T\mathbf{R} + \lambda \mathbf{I})^{-1}$, $\lambda$ is a hyperparameter for regularization, and $\oslash$ denotes the element-wise division.
    \item In the case of $\infty$-AE~\cite{sachdeva2022infinite}, $\mathbf{K} \in \mathbb{R}^{|U| \times |U|}$ is a Gram matrix that represents the similarity of users, i.e., $\mathbf{K}_{u,v} \coloneq \mathbb{K}(\mathbf{R}_u, \mathbf{R}_v)$ for all $u,v \in U$. $\mathbb{K}$ denotes a NTK, $\mathbb{K}: \mathbb{R}^{|I|} \times \mathbb{R}^{|I|} \mapsto \mathbb{R}$ over an Reproducing Kernel Hilbert Space $\mathcal{H}$ of a single-layer autoencoder with an activation function $\sigma$. A closed-form solution is given by $\mathbb{K}(\mathbf{R}_u, \mathbf{R}_v) = \tilde{\sigma}(\mathbf{R}_u^T\mathbf{R}_v) + \tilde{\sigma}'(\mathbf{R}_u^T\mathbf{R}_v) \cdot \mathbf{R}_u^T\mathbf{R}_v$. $\tilde{\sigma}$ represents the dual activation of $\sigma$ and $\tilde{\sigma}'$ represents its derivative. $\lambda$ is a regularization parameter.
    \item In the case of our SVD-AE, which will be described in detail in Section~\ref{sec:svd-ae}, $\tilde{\mathbf{R}}$ is a normalized rating matrix given by $\tilde{\mathbf{R}}=\mathbf{D}_U^{-\frac{1}{2}}\mathbf{R}\mathbf{D}_I^{-\frac{1}{2}}$. $\mathbf{D}_U = \text{diagMat}(\mathbf{R} \cdot \mathbf{1})$ and $\mathbf{D}_I = \text{diagMat}(\mathbf{1}^T\mathbf{R})$ are degree matrices. $\mathbf{Q} \in \mathbb{R}^{|U| \times m}$ and $\mathbf{V} \in \mathbb{R}^{|I| \times m}$ are the top-$m$ singular vectors, and $\tilde{\mathbf{\Sigma}}$ denotes the top-$m$ singular values of $\tilde{\mathbf{R}}$.
\end{enumerate}

For all methods, \emph{linear regression} is used to discover the closed-form solution of learning the optimal $\hat{\mathbf{R}}$. We show how two simple regression methods --- ridge regression and kernelized ridge regression --- solve the problem in the closed-form in the following.

\paragraph{Ridge Regression.}
EASE applies ordinary least squares regression and derives $\mathbf{B}$ using the following convex objective:
\begin{align}
        \operatorname*{min}_{\mathbf{B}}& \quad \| \mathbf{R} - \mathbf{R}\mathbf{B} \|_F^2 + \lambda \cdot \|\mathbf{B}\|_F^2, \label{ease:1} \\
        \textrm{s.t. }& \quad \text{diag}(\mathbf{B}) = 0, \label{ease:2}
\end{align} where $\| \cdot \|_F$ denotes the Frobenius norm and $\lambda$ is the regularization parameter. The problem in Eq.~\eqref{ease:1} is a simple ridge regression and the solution is well-known as follows:

\begin{equation}
    \mathbf{B} = (\mathbf{R}^T\mathbf{R} + \lambda \mathbf{I})^{-1} \mathbf{R}^T\mathbf{R}.
\end{equation}

EASE also includes Lagrangian multipliers into the problem to fulfill the zero diagonal constraints in Eq.~\eqref{ease:2}. This is important in order to avoid trivial solution $\mathbf{B} = \mathbf{I}$. 

\paragraph{Kerenelized Ridge Regression.}
By using the estimated NTK, $\infty$-AE conducts Kernelized Ridge Regression (KRR) which is the same as training an infinitely-wide autoencoder for an infinite number of SGD steps. They reduce the problem to KRR as follows: 

\begin{align}
    \operatorname*{arg min}_{[\alpha_j]_{j=1}^{|U|}}& \quad \sum_{u \in U} \|\mathbf{R}_u - f(\mathbf{R}_u|\alpha) \|_2^2 + \lambda \cdot \|f\|_\mathcal{H}^2, \label{inf:1} \\
    \textrm{s.t. }& \quad f(\mathbf{R}_u | \alpha) = \sum_{j=1}^{|U|} \alpha_j \cdot \mathbb{K}(\mathbf{R}_u, \mathbf{R}_{u_j}) , \label{inf:2}
\end{align} where $\alpha \coloneqq [\alpha_1;\alpha_2 ...; \alpha_{|U|}] \in \mathbb{R}^{|U| \times |I|}$ are the set of dual parameters to estimate and $\lambda$ is a regularization constant. The KRR problem in Eq.~\eqref{inf:1} has a closed-form solution given by $\hat{\alpha} = (\mathbf{K} + \lambda \mathbf{I})^{-1} \cdot \mathbf{R}$ s.t. $\mathbf{K}_{u,v} \coloneq \mathbb{K}(\mathbf{R}_u, \mathbf{R}_v)$ $\forall u,v$.

\section{SVD-AE}\label{sec:svd-ae}
In this section, we describe our three design goals, followed by our detailed method design and its low-rank approximation. Finally, we discuss the advantages of SVD-AE and explore how it relates to other recommendation methods.

\paragraph{Design Goals.} Our proposed problem formulation has been designed for the following three design goals:
\begin{enumerate}
    \item \textbf{(Robustness)} Its problem formulation should include an effective inductive bias that can lead to high noise robustness;
    \item \textbf{(Complexity)} Its closed-form solution should be no more complicated than existing methods based on different closed-form solutions, e.g., EASE and $\infty$-AE;
    \item \textbf{(Accuracy)} Its recommendation accuracy ought to be comparable to that of  existing CF methods.
\end{enumerate}

\paragraph{Method Design.} Our SVD-AE solves yet another linear regression problem for the recommendation --- however, its problem formulation can still be in the generalized linear regression regime of Eq.~\eqref{eq:generalized_ae}, i.e., 
\begin{equation}
    \operatorname*{min}_\mathbf{B} \quad \|\mathbf{R}-\tilde{\mathbf{R}}\mathbf{B}\|^2,
\end{equation}where $\tilde{\mathbf{R}}$ is a normalized adjacency matrix given by $\tilde{\mathbf{R}}=\mathbf{D}_U^{-\frac{1}{2}}\mathbf{R}\mathbf{D}_I^{-\frac{1}{2}}$. $\mathbf{D}_U = \text{diagMat}(\mathbf{R} \cdot \mathbf{1})$ and $\mathbf{D}_I = \text{diagMat}(\mathbf{1}^T\mathbf{R})$ are degree matrices. We discuss in the following subsection that our problem formulation satisfies the three design goals. 

Based on Theorem~\ref{thm:1} and \ref{thm:2}, we can derive the closed-form solution of $\mathbf{B}$ as follows:
\begin{equation}
    \mathbf{B} =  \tilde{\mathbf{R}}^{+}\mathbf{R} = \mathbf{V} \mathbf{\Sigma}^{+}\mathbf{Q}^T\mathbf{R}.
\end{equation}

\begin{theorem}\label{thm:1}
    The least squares solutions of the minimum norm of the linear system $\tilde{\mathbf{R}}\mathbf{B} = \mathbf{R}$ is given by
        \begin{equation}
            \mathbf{B} = \tilde{\mathbf{R}}^{+}\mathbf{R},
        \end{equation}where $\tilde{\mathbf{R}}^{+}$ is the the pseudo-inverse of $\tilde{\mathbf{R}}$. 
\end{theorem}

\begin{theorem}\label{thm:2} 
    Let $\tilde{\mathbf{R}}$ be a normalized adjacency matrix and the SVD of $\tilde{\mathbf{R}}$ be
        \begin{equation}
            \tilde{\mathbf{R}} = \mathbf{Q}  \mathbf{\Sigma} \mathbf{V}^T,
        \end{equation}where $\mathbf{Q}$, $\mathbf{V}$ are both orthogonal matrices and $\mathbf{\Sigma}=\mathrm{diagMat}(\sigma_1, ..., \sigma_M)$ is a diagonal matrix containing the singular values of $\tilde{\mathbf{R}}$. Then the pseudo-inverse of $\tilde{\mathbf{R}}$ denoted by $\tilde{\mathbf{R}}^{+}$ is defined as:
    \begin{equation}
        \tilde{\mathbf{R}}^{+} = \mathbf{V} \mathbf{\Sigma}^{+} \mathbf{Q}^T,
    \end{equation}where $\mathbf{\Sigma}^{+}$ is given by $\mathbf{\Sigma}^{+}=\text{diagMat}(1/{\sigma_1}, ..., 1/{\sigma_M})$.
\end{theorem}

\paragraph{Low-rank Inductive Bias.} However, it is known that the pseudo-inverse can provide serious effects when used in the presence of noise~\cite{stiles1985effect,murakami1987improvement}. The truncated SVD is one method for stabilizing the pseudo-inverse among others.

The truncated SVD uses only the first $m$ columns of $\mathbf{Q}$ and $\mathbf{V}$ and the first $m$ largest singular values. The relatively large and small singular values of the matrix with rank $M$ typically cluster into two groups, i.e.,
\begin{equation}
    \underbrace{\sigma_1 \geq \sigma_2 \geq ... \geq \sigma_m}_\text{large group} \gg \underbrace{\sigma_{m+1} \geq ... \geq \sigma_M}_\text{small group}.
\end{equation}

Noise reduction can be executed by utilizing only the significant components represented by $m$ singular values from the larger group, considering the smaller group as noise. Truncated SVD is widely used in recommender systems to handle noise in the rating matrix $\mathbf{R}$~\cite{shen2021powerful,peng2022svd}. Since the rating matrix is typically vast and sparse, it is also useful for speeding up calculations and information retrieval.

We use the normalized adjacency matrix whose singular values are between 0 and 1 as shown in the next theorem. This avoids having singular values that are too small or large.

\begin{theorem}\label{thm:3}
        Let $\sigma_1 \geq \sigma_2 \geq ... \geq \sigma_M$ be the singular values of $\tilde{\mathbf{R}}$ given by $\tilde{\mathbf{R}}=\mathbf{D}_U^{-\frac{1}{2}}\mathbf{R}\mathbf{D}_I^{-\frac{1}{2}}$, then
        \begin{equation}
            0 \leq \sigma_M \le... \leq \sigma_2 \leq \sigma_1  \leq 1.
        \end{equation}
\end{theorem}

Using the truncated SVD, $\mathbf{Q} \in \mathbb{R}^{|U| \times m}$ and $\mathbf{V} \in \mathbb{R}^{|I| \times m}$ are the top-$m$ singular vectors and $\tilde{\mathbf{\Sigma}}$ denotes the top-$m$ singular values of $\tilde{\mathbf{R}}$ given by $\tilde{\mathbf{\Sigma}}=\text{diagMat}(\sigma_1,...,\sigma_m)$. Thus, the following is the reconstructed (or inferred) interaction matrix $\hat{\mathbf{R}}$ for recommendation:
\begin{equation}
    \hat{\mathbf{R}} = \tilde{\mathbf{R}}\mathbf{B} =  \tilde{\mathbf{R}} \cdot \mathbf{V} \tilde{\mathbf{\Sigma}}^{+}\mathbf{Q}^T\mathbf{R}.
\end{equation}

\underline{Proofs of the theorems can be found in the \emph{Appendix~\ref{app:proof}}.}

\begin{figure}[t]
    \centering
    \includegraphics[width=0.88\columnwidth]{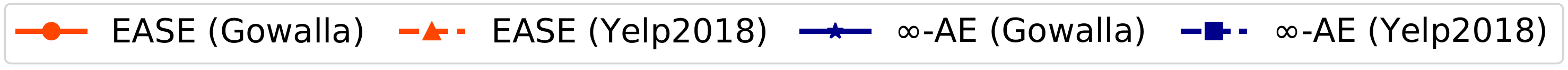} 
    \subfigure[MSE]{\includegraphics[width=0.44\columnwidth]{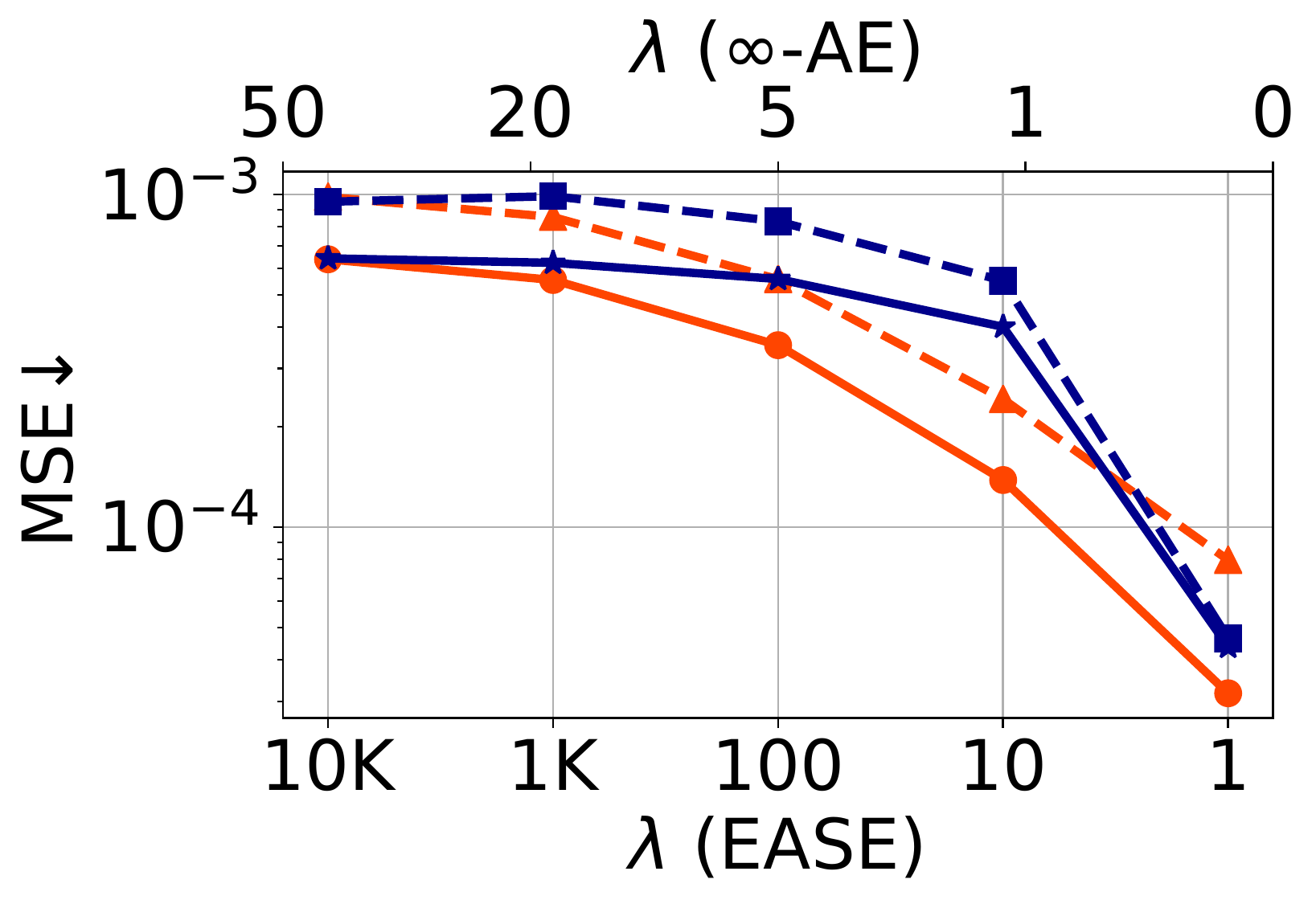}}
    \subfigure[HR@10]{\includegraphics[width=0.44\columnwidth]{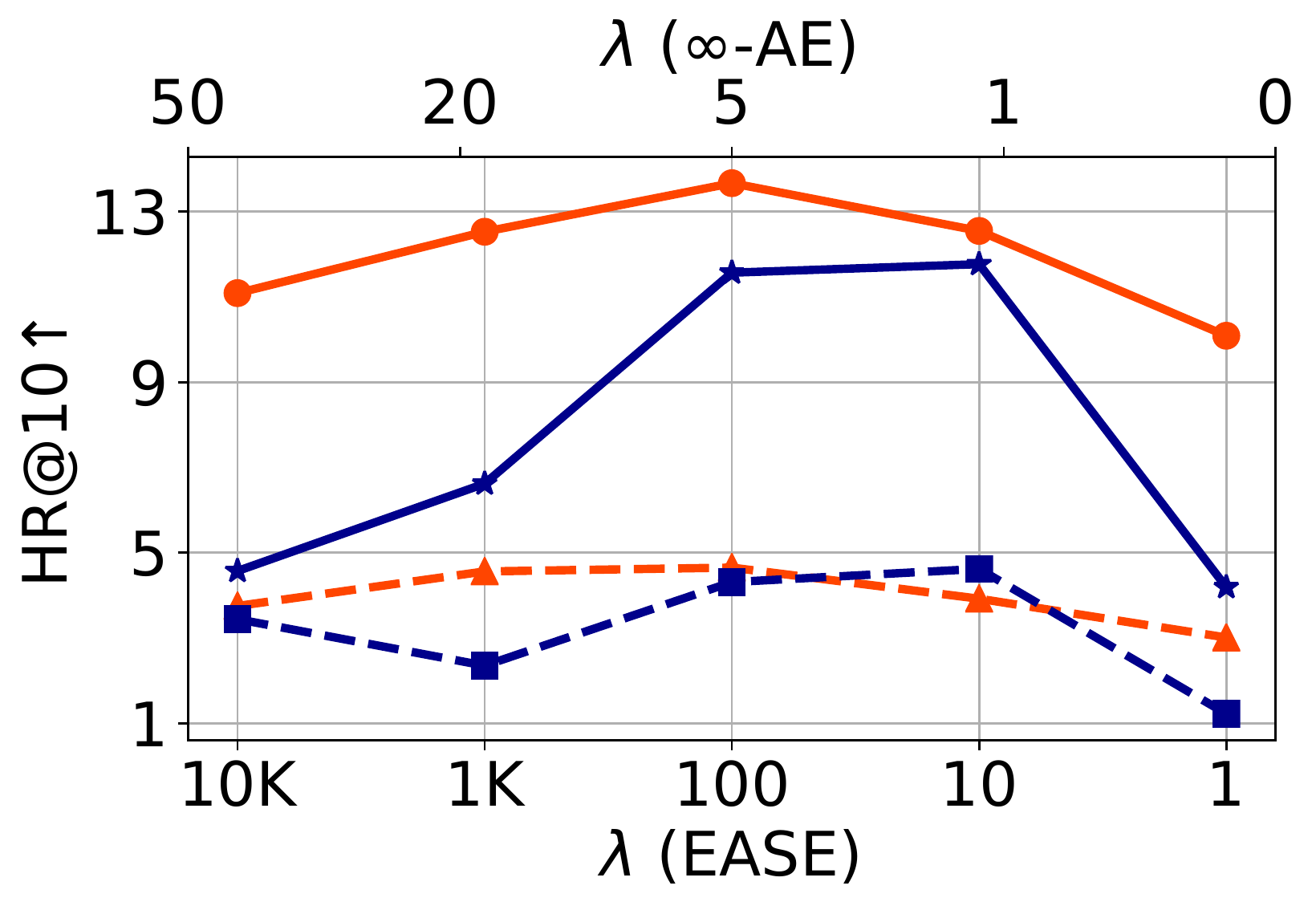}}
    \caption{The performance comparison with different regularization parameters. The \boldsymbol{$y$}-axis of (a) is on a log scale.}
    \label{fig:overfitting_baselines}
\end{figure}

\subsection{Why SVD-AE Achieves the Best Balance?}\label{sec:discussion}
\paragraph{The Presence of Noise.}
Many algorithms used in recommender systems utilize observed ratings to predict unknown ratings. The user ratings can be seen as the ground truth of the user's preference, which is an underlying assumption in this method. However, given the presence of noisy data, inferring unknown ratings is challenging. A user-item interaction record does not always mean that users are satisfied with consumed items, which frequently happens for implicit interactions. Users are also inconsistent in giving their feedback, creating an undetermined amount of noise that makes it more challenging to infer unreported ratings~\cite{amatriain2009rate,sar2015data,said2012users,amatriain2009like}. Therefore, the two types of interaction noises can be summarized as follows:
\begin{enumerate}
\item (Noisy 1 in $\mathbf{R}$) A user is not satisfied with an item, but an interaction is observed between them;
\item (Noisy 0 in $\mathbf{R}$) A user will be satisfied with an item, but there are no observed interactions between them. The ultimate goal of recommendation is to accurately reveal these hidden interactions.
\end{enumerate}

Here, one can argue that EASE and $\infty$-AE address these noises by preventing overfitting through the regularization parameter $\lambda$. 
In Figure~\ref{fig:overfitting_baselines}, we show that minimizing the Mean Squared Error (MSE) loss in Eq.~\eqref{eq:generalized_ae} does not guarantee the recommendation performance. We can observe that performance does not increase when EASE and $\infty$-AE use smaller $\lambda$ to get closer to $\mathbf{R}$, a trivial solution of $\hat{\mathbf{R}}$. For all cases, the performance is at the lowest even when the smallest $\lambda$ is applied. Using the appropriate $\lambda$ leads to higher performance by avoiding overfitting the noise-inherent original rating matrix.

However, rather than using a regularization parameter, we adopt the truncated SVD to strongly and directly prevent overfitting of noisy interactions. In the perspective of \emph{signal processing}~\cite{ortega2018graph}, the above two types of noises correspond to \emph{high-frequency signals} in $\mathbf{R}$. In other words, they typically make local abrupt changes in the adjacency structure of $\mathbf{R}$. For instance, the second type means that a hidden element (edge) of $\mathbf{R}$ is zero when its incident elements (edges) frequently have 1 --- the first type can also be interpreted similarly. One effective noise removal method is to smooth out those local large changes (i.e., high-frequency signals) by utilizing only the $m$ lowest frequency signals. To this end, our design is based on the effective low-rank inductive bias based on the truncated SVD with a rank parameter of $m$. Among various possible inductive biases, we have chosen the (low-rank) truncated SVD for its effectiveness.

\begin{figure}[t]
    \centering
    \subfigure[$\mathbf{R}$]{\includegraphics[width=0.31\columnwidth]{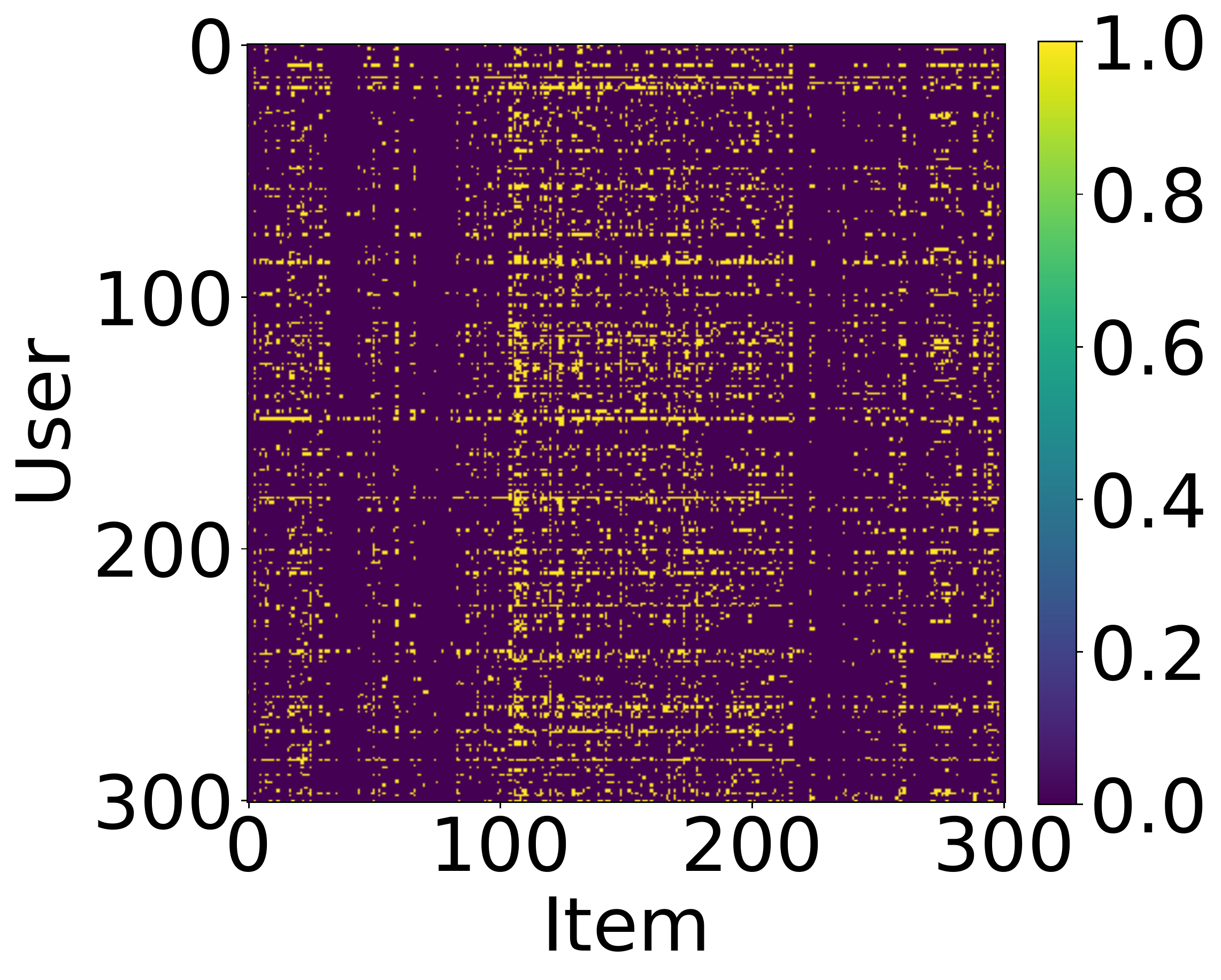}}
    \subfigure[$\tilde{\mathbf{R}}$]{\includegraphics[width=0.31\columnwidth]{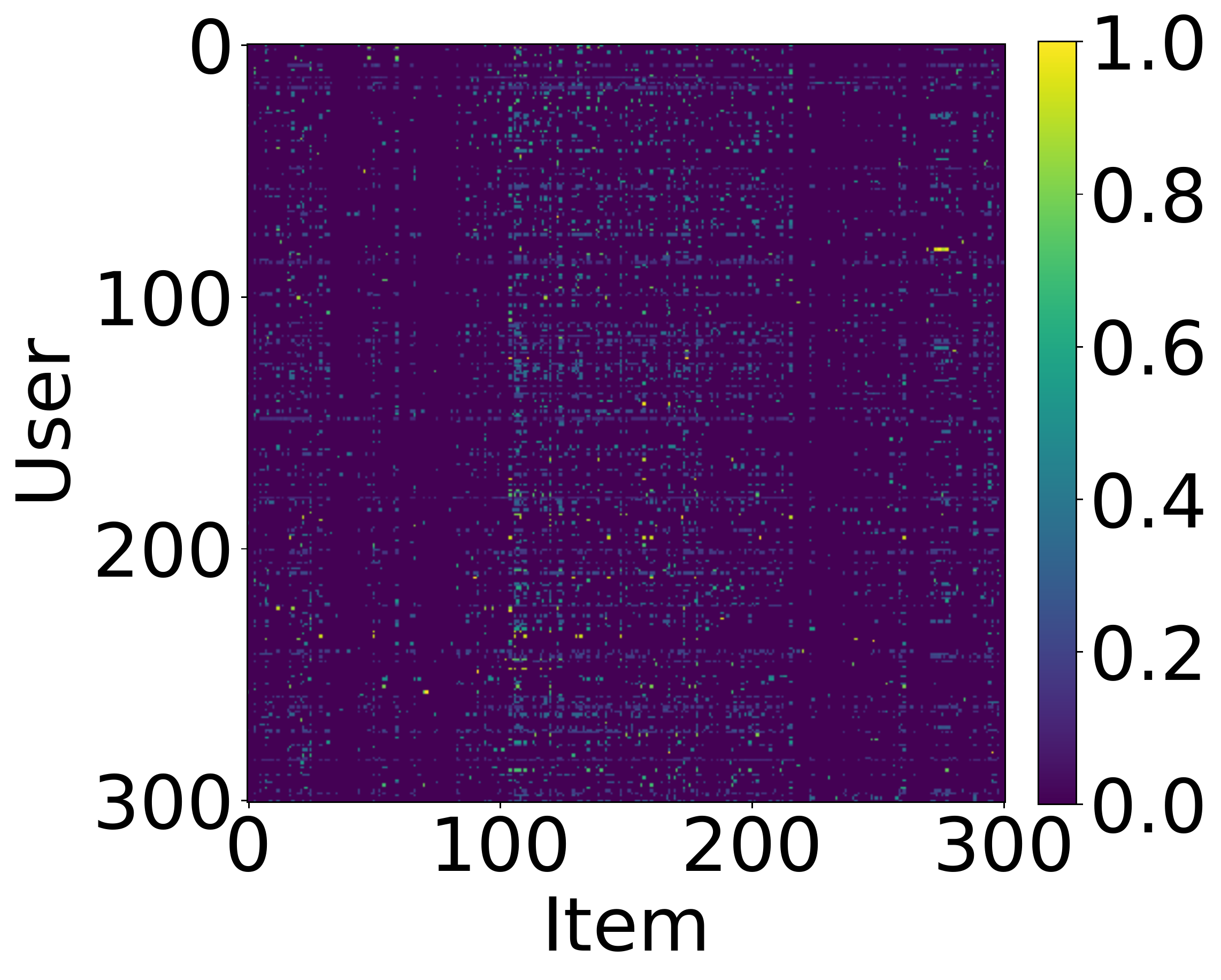}}
    \subfigure[$\mathbf{Q}\tilde{\mathbf{\Sigma}}\mathbf{V}^T$ ]{\includegraphics[width=0.32\columnwidth]{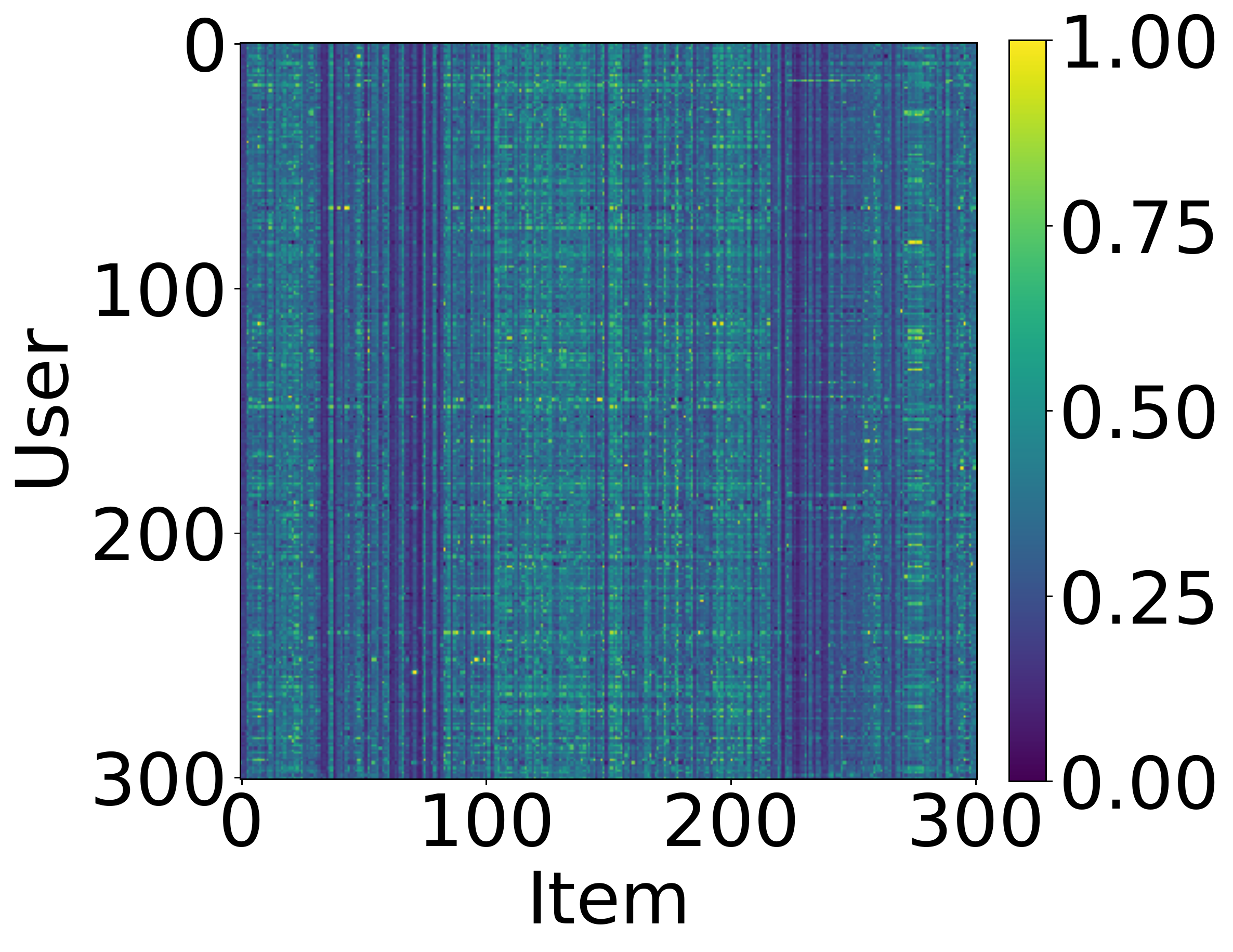}}
    \subfigure[The histogram of $\tilde{\mathbf{R}}$]{\includegraphics[width=0.49\columnwidth]{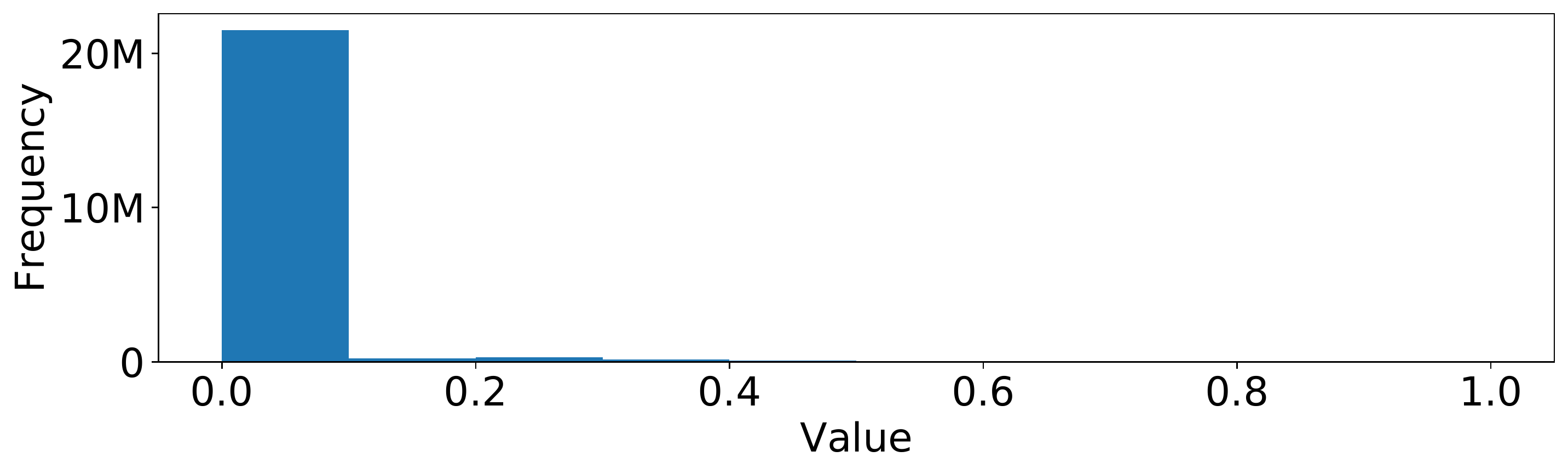}}
    \subfigure[The histogram of $\mathbf{Q}\tilde{\mathbf{\Sigma}}\mathbf{V}^T$ ]{\includegraphics[width=0.49\columnwidth]{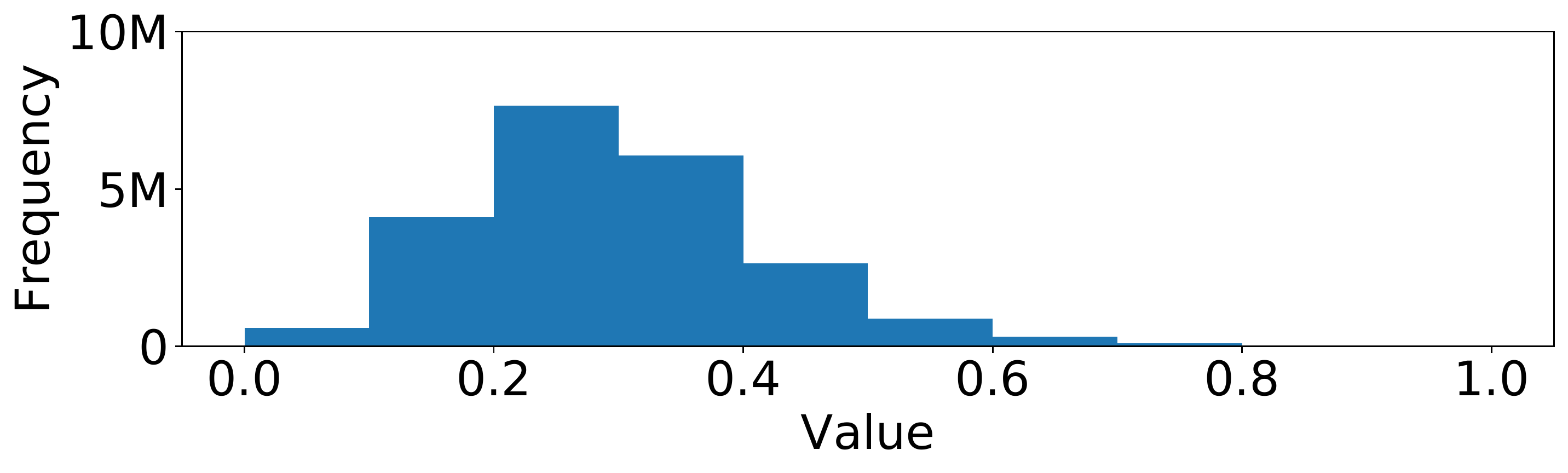}}
    \caption{The smoothing effect of the truncated SVD in reducing noise. All values are normalized between 0 and 1 for better representation.
    For (a)-(c), 300 users and items from ML-1M are sampled and all interactions of ML-1M are counted for (d) and (e).}
    \label{fig:trunc_svd}
\end{figure}

Figures~\ref{fig:trunc_svd} (a), (b), and (c) show the smoothing effect of the truncated SVD. We discover that stable structured data extraction is possible by reconstructing the adjacency matrix using the truncated SVD. The truncated SVD converts the original data into the good representation by reducing noise. 
Comparing Figures~\ref{fig:trunc_svd} (d) and (e), it is clear that
the reconstructed representation (i.e., Figure~\ref{fig:trunc_svd} (e)) has a range of values as opposed to the existing data distribution (i.e., Figure~\ref{fig:trunc_svd} (d)), which had only values very close to zero.


\paragraph{Computational Cost.}
In this part, we examine the computational cost of SVD-AE. We use a fast randomized truncated SVD~\cite{halko2011finding} and its computational complexity is $\mathcal{O}(|U| \cdot |I| \cdot log(m)+(|U|+|I|) \cdot m^2)$. This can be done in the pre-processing step --- there exist even more lightweight methods to approximate the truncated SVD~\cite{shishkin2019fast,feng2018faster}, and one can try to further decrease the time complexity by using them. In the case of EASE and $\infty$-AE, they need to compute the matrix inversion of the data Gram matrix, $\mathbf{R}^T\mathbf{R} \in \mathbb{R}^{|I| \times |I|}$ or $\mathbf{R}\mathbf{R}^T \in \mathbb{R}^{|U| \times |U|}$. Then the computation complexity would be $\mathcal{O}(|I|^{2.376})$ or $\mathcal{O}(|U|^{2.376})$ that are computationally-expensive. However, since we do not have to compute the matrix inversion, the computational cost is significantly reduced. The overall computation cost comes out to be $\mathcal{O}(|U| \cdot |I|^2)$. The computation complexity of EASE is $\mathcal{O}(|I| \cdot (|I|-1)^{2.376})$ with the additional cost $\mathcal{O}(|U| \cdot |I|^2)$ for pre-processing step whereas $\infty$-AE's cost is $\mathcal{O}(|U|^2 \cdot |I| + |U|^{2.376})$. 


\begin{table*}[t]
\small
\setlength{\tabcolsep}{3pt}
\centering
\begin{tabular}{l l rrrrrrrr r}
\toprule
Dataset & Measure   & MF-BPR & NeuMF & NGCF & LightGCN    & GF-CF & MultVAE & EASE & $\infty$-AE    & \textbf{SVD-AE} \\ \midrule
\multirow{5}{*}{Gowalla}  & HR@10 & 12.08  & 8.87  & 11.70 & 14.00 & {\ul 14.08} & 11.88 & 13.67 & 11.77 & \textbf{14.40}  \\
 & HR@100   & 32.84  & 27.25 & 32.22 & {\ul 37.40} & \textbf{38.84} & 33.56 & 35.74 & 34.20 & 37.34 \\
 & NDCG@10  & 12.09  & 8.27  & 11.70 & {\ul 13.77} & 13.50          & 11.30 & 13.15       & 10.84          & \textbf{13.94}  \\
                          & NDCG@100 & 18.44  & 13.96 & 17.95                      & 21.04       & \textbf{21.25} & 18.11      & 20.08       & 17.97          & {\ul 21.15}     \\
                          & PSP@10   & 1.92   & 1.37  & 1.76                       & 2.26        & {\ul 2.47}     & 2.09       & 2.31        & 2.02           & \textbf{2.48}   \\ \midrule
\multirow{5}{*}{Yelp2018} & HR@10    & 3.86   & 3.25  & 3.68                       & 4.32        & {\ul 4.87}     & 4.31       & 4.65        & 4.62           & \textbf{4.90}   \\
                          & HR@100   & 16.97  & 14.39 & 16.88                      & 19.01       & \textbf{20.86} & 18.75      & 17.74       & 18.33          & {\ul 19.79}     \\
                          & NDCG@10  & 3.70   & 3.08  & 3.50                       & 4.19        & {\ul 4.66}     & 4.10       & 4.55        & 4.48           & \textbf{4.74}   \\
                          & NDCG@100 & 8.50   & 7.16  & 8.33                       & 9.57        & \textbf{10.53} & 9.37       & 9.37        & 9.54           & {\ul 10.22}     \\
                          & PSP@10   & 0.34   & 0.27  & 0.32                       & 0.39        & {\ul 0.44}     & 0.43       & 0.42        & 0.43           & \textbf{0.45}   \\ \midrule
\multirow{5}{*}{ML-1M}    & HR@10    & 28.49  & 27.95 & 28.40                      & 29.07       & 30.81          & 27.86      & 30.43       & {\ul 31.15}    & \textbf{31.79}  \\
                          & HR@100   & 57.21  & 54.24 & 57.58                      & 57.62       & 59.10          & 57.67      & 57.74       & \textbf{60.75} & {\ul 59.33}     \\
                          & NDCG@10  & 29.84  & 29.36 & 29.43                      & 30.30       & {\ul 32.37}    & 28.44      & 31.90       & 32.27          & \textbf{33.55}  \\
                          & NDCG@100 & 39.47  & 37.98 & 39.27                      & 39.95       & 42.00          & 39.34      & 40.95       & {\ul 42.54}    & \textbf{42.57}  \\
                          & PSP@10   & 2.89   & 2.75  & 2.90                       & 3.01        & {\ul 3.17}     & 3.13       & 3.16        & \textbf{3.22}  & \textbf{3.22}   \\ \midrule
\multirow{5}{*}{ML-10M}   & HR@10    & 33.28  & 25.75 & \multirow{5}{*}{Timed Out} & 34.79       & 35.10          & 34.20      & {\ul 36.30} & 35.83          & \textbf{36.76}  \\
                          & HR@100   & 64.16  & 58.63 &                            & 64.11       & 64.23          & 64.55      & {\ul 64.78} & 64.48          & \textbf{64.80}  \\
                          & NDCG@10  & 33.58  & 25.32 &                            & 35.60       & 36.02          & 34.48      & {\ul 37.63} & 36.93          & \textbf{37.75}  \\
                          & NDCG@100 & 44.32  & 36.90 &                            & 46.14       & 45.71          & 45.23      & {\ul 46.74} & 46.27          & \textbf{46.97}  \\
                          & PSP@10   & 4.60   & 3.52  &                            & 4.69        & 4.73           & {\ul 4.82} & 4.76        & 4.74           & \textbf{4.93}   \\ \bottomrule
\end{tabular}
\caption{Performance evaluation of overall performance among SVD-AE and baselines. \textbf{Bold} values indicate the best values in each row, while \underline{underlined} values indicate the second-best values. Higher values are preferable for all measures. }
\label{tbl:main_exp}
\end{table*}

\subsection{Relation to Other Methods}\label{sec:relation}


Table~\ref{tbl:cmp1} summarizes the key distinctions between our SVD-AE and current lightweight methods. We will now explore their relationships and compare SVD-AE to others. 

EASE, $\infty$-AE, and SVD-AE share the CF paradigm, utilizing a shallow autoencoder-based linear regression for problem resolution, leading to closed-form optimization. In the case of EASE and $\infty$-AE, they are based on ridge regression with the regularization parameter $\lambda$ to avoid overfitting. Unlike previous methods, SVD-AE does not require an additional regularizer by directly using the truncated SVD as a low-rank inductive bias, to avoid overfitting more sophisticated. As discussed earlier in \textit{The Presence of Noise}, this design choice is based on the smoothing effect and the efficient low-rank inductive bias of truncated SVD. For $\infty$-AE, it needs to build a neural network to use NTK, but other methods can reconstruct the matrix using a simple matrix computation. This complexity results in $\infty$-AE having longer execution times with large datasets than EASE or SVD-AE. This limitation will be discussed in more detail in a later section.

GF-CF~\cite{shen2021powerful} is a graph filter-based model, which is a combination of the linear graph filter and the ideal low-pass graph filter. 
In GF-CF, the truncated SVD is used to construct an ideal low-pass graph filter rather than to prevent overfitting.
In other words, it is used as an auxiliary to add information to the linear graph filter, rather than as the main mechanism. 
However, in SVD-AE, the information obtained solely through the truncated SVD is directly used as the main objective. 
This design choice not only simplifies the process but also enhances efficiency, as it employs an autoencoder and involves only a linear regression problem.
It is in contrast to the graph-based approach, such as GF-CF, which typically requires more complex designs and computational steps.

Additionally, comparing low-rank and full-rank methods is another intriguing aspect of our research. SVD-AE is designed on top of our low-rank inductive bias, while GF-CF also partially utilizes it. This contrasts EASE and $\infty$-AE, which use full-rank information. Moreover, $\infty$-AE is an over-parameterized model with an infinitely wide layer. Our experiments will show that low-rank methods, due to their low-rank inductive bias, outperform full-rank and over-parametrized methods for sparse data. 

Finally, using a Gram matrix is another distinctive feature of these approaches. In recommender systems, it is common to compute an item-item or user-user similarity matrix from which to derive a new user-item interaction matrix. EASE and GF-CF utilize an item-item matrix, while $\infty$-AE uses a user-user matrix. SVD-AE does not use any of them but directly infers user-item interactions, which reduces the overall computation cost as shown in Section~\ref{sec:discussion}.

\section{Experiments}
In this section, we describe the experimental settings and then compare our SVD-AE with state-of-the-art CF methods. 
We also discuss our study of the robustness of various algorithms. 

\subsection{Experimental Environments}
\paragraph{Datasets.}
In our experiments, we use four of the most frequently used publicly available datasets~\cite{cho2011friendship,harper2015movielens}: Gowalla, Yelp2018, ML-1M, and ML-10M. 
See \emph{Appendix}~\ref{app:dataset} for details on datasets.


\begin{table}[t]
\small
\setlength{\tabcolsep}{2.3pt}
\centering
\begin{tabular}{l rrrr}
\toprule
\multirow{2}{*}{Model} & \multicolumn{2}{c}{ML-1M} & \multicolumn{2}{c}{ML-10M} \\ \cmidrule(r){2-3} \cmidrule(r){4-5}
                       & Pre-processing  & Training & Pre-processing  & Training \\ \midrule
MF-BPR                 & N/A             & 24.48m   & N/A             & 48.56h   \\
NeuMF                  & N/A             & 2.68h    & N/A             & 145.98h  \\
NGCF                   & N/A             & 4.83h    & N/A             & Timed Out\\
LightGCN               & N/A             & 2.44h    & N/A             & 132.97h  \\
GF-CF                  & 4.62s           & 6.37s    & 28.98s          & 1260.80s \\
MultVAE                & N/A             & 221.78s  & N/A             & 1293.76s \\
EASE                   & 4.52s           & 5.72s    & 52.63s          & 6.05s    \\
$\infty$-AE            & N/A             & 2.24s    & N/A             & 388.39s  \\ \midrule
\textbf{SVD-AE}        & 0.54s           & 2.06s    & 47.59s          & 3.06s    \\ \bottomrule
\end{tabular}
\caption{Efficiency comparison on overall computation time.} 
\label{tbl:computation}
\end{table}

\paragraph{Evalutation Protocol.}
We compare SVD-AE with 8 baselines: MF-BPR~\cite{rendle2012bpr}, NeuMF~\cite{he2017neural}, NGCF~\cite{wang2019neural}, LightGCN~\cite{he2020lightgcn}, GF-CF, MultVAE~\cite{liang2018variational}, EASE, and $\infty$-AE.
We provide details on baselines and their hyperparameter configurations in \emph{Appendix}~\ref{app:baseline}.
For a fair comparison with previous studies, we use the same train/validation/test splits as done in ~\cite{he2020lightgcn,shen2021powerful,sachdeva2022infinite}.
We evaluate the accuracy of all of the methods using HR@k, NDCG@k, and PSP@k~\cite{jain2016extreme}.

\paragraph{Implementation Details.}
We need to take $m$ as a rank parameter when running the truncated SVD. For simplicity, we use $m = \lfloor \gamma \times \text{min}(|U|, |I|)\rceil$, where $\lfloor ... \rceil$ function denotes the rounding function. We search for $\gamma$ in the range of $[0.01, 0.02, 0.03, 0.04, 0.05 ]$ with the validation set with HR@10. 
Note that \underline{$\gamma$ is our sole hyperparameter}, and $\gamma = 0.04$ is universally optimal for all datasets, simplifying the selection of suitable rank parameter $m$.

\subsection{Performance Comparison}
The performance of the proposed method and other benchmark models are shown in Table~\ref{tbl:main_exp}. Overall, our SVD-AE achieves comparable to or better performance than other various methods. SVD-AE yields the best performance for HR@10, NDCG@10, and PSP@10 across all datasets. Among all baselines, GF-CF performs well on the relatively sparse datasets, Gowalla and Yelp2018. AE-based models such as EASE and $\infty$-AE perform rather poorly for these datasets. SVD-AE performs well on sparse datasets compared to other AE-based models, and outperforms EASE and $\infty$-AE on Gowalla by 5.3\% and 22.3\%, respectively, in terms of HR@10. In less sparse datasets like ML-1M and ML-10M, where user-item interactions are comparatively sufficient, AE models outperform non-AE models. $\infty$-AE and EASE show good performance on ML-1M and ML-10M. But once again, SVD-AE has the greatest performance aside from HR@100 on ML-1M. In highly sparse datasets, AE-based models struggle to reconstruct the rating matrix for deriving the inferred matrix. However, SVD-AE maintains consistent performance across dataset types, possibly owing to our low-rank inductive bias.

One can argue that the accuracy improvements of SVD-AE over baselines are often incremental. However, we emphasize that the goal of this work is to design \emph{a method that achieves a reasonable level of accuracy compared to baselines, while considerably diminishing computational time and improving robustness to noisy interactions}, which will be validated in the subsequent sections.

\subsection{Efficiency Comparison}

We contrast the training 
time for our method with that of baseline models. For all models, we employ the same experimental setups and batch sizes. For runtime studies, we use the smallest and largest datasets, ML-1M and ML-10M. 

Table~\ref{tbl:computation} illustrates the huge discrepancy between the two methods—training models (i.e., MF-BPR, NeuMF, NGCF, LightGCN, and MultVAE) and non-training models (i.e., EASE, GF-CF, $\infty$-AE, and SVD-AE).
The limitation of training models is clearly seen when the data size is huge. 
Notably, their efficiency may be acceptable for small datasets, but may take several days to train for large datasets.
Especially in the case of NGCF, it is impracticable to train the model for ML-10M.
Among the training models, MultVAE, the AE-based model shows a tolerable calculation time regardless of the size of the dataset. 

The non-training models like GF-CF, EASE, and $\infty$-AE have relatively faster computation times. The presence of closed-form solutions makes this feasible. In the case of $\infty$-AE, however, compared to other AE-based models, it takes a lot longer to compute as the size of the data increased. 
This is because $\infty$-AE employs the NTK, making it more complex than lighter methods. Most importantly, our SVD-AE can achieve good performance in less time, which further illustrates the high efficiency of SVD-AE.

We also report pre-processing time as some methods including our method require pre-processing. GF-CF and SVD-AE require the truncated SVD before training. In the case of EASE, it computes $\mathbf{R}^T\mathbf{R}$ in the pre-processing step. Table~\ref{tbl:computation} shows that the time taken for the truncated SVD increases with the data size. However, since 10M interactions take only a minute, the procedure is not onerous.


\begin{figure}[t]
    \centering
    \subfigure[MSE (Gowalla)]{\includegraphics[width=0.48\columnwidth]{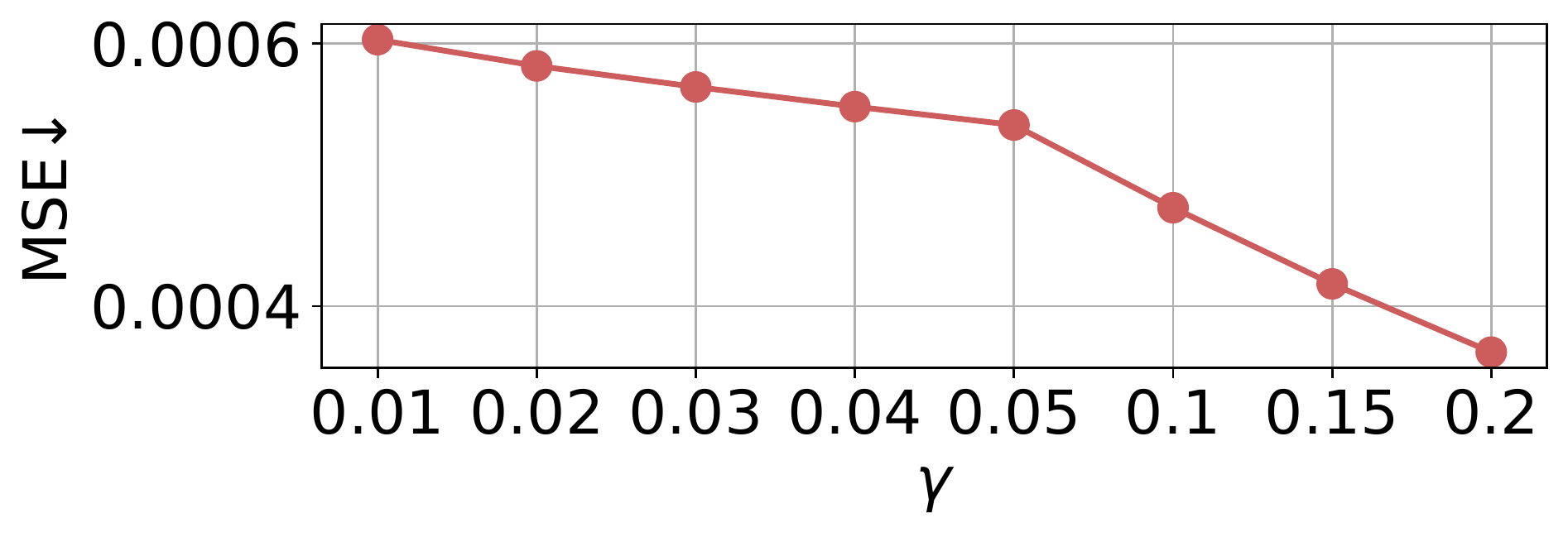}}
    \subfigure[Accuracy (Gowalla)]{\includegraphics[width=0.48\columnwidth]{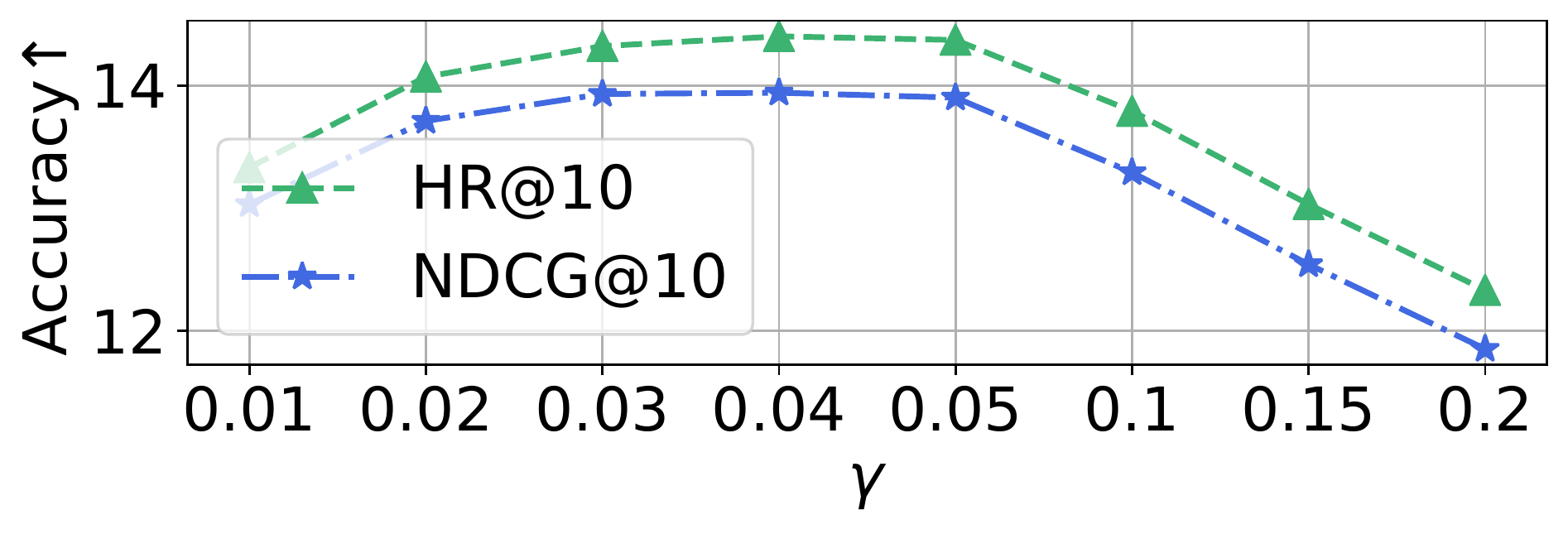}}
    \subfigure[MSE (Yelp2018)]{\includegraphics[width=0.48\columnwidth]{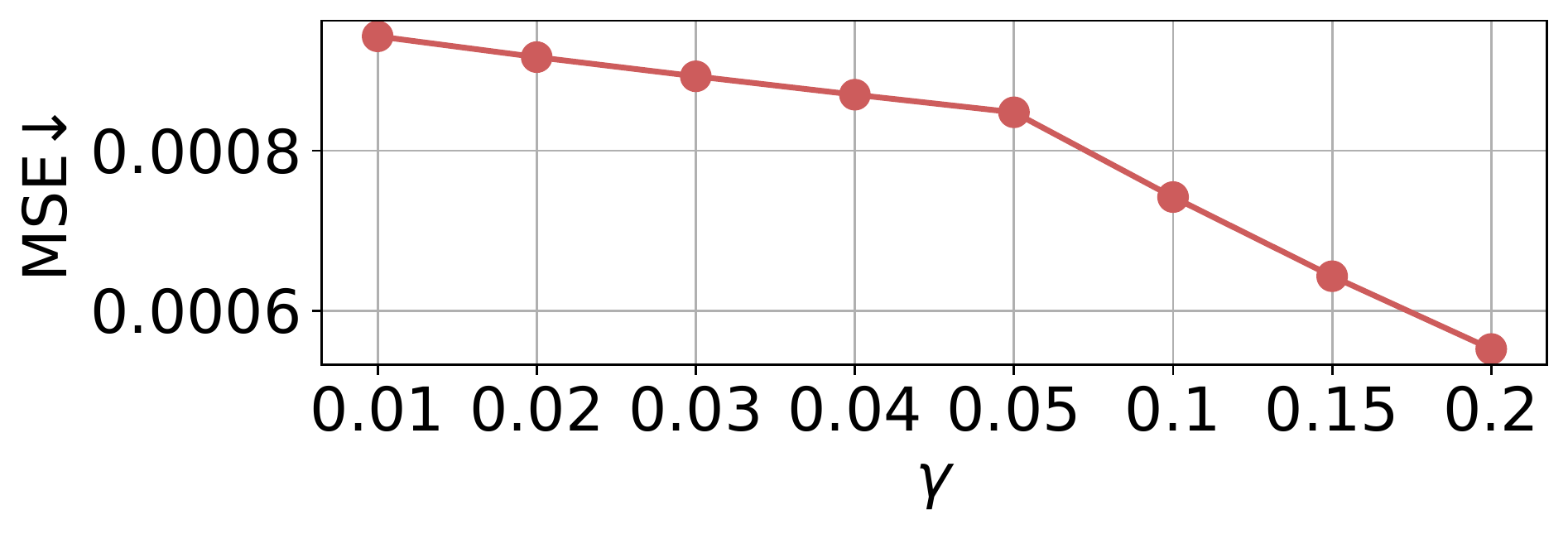}}
    \subfigure[Accuracy (Yelp2018)]{\includegraphics[width=0.48\columnwidth]{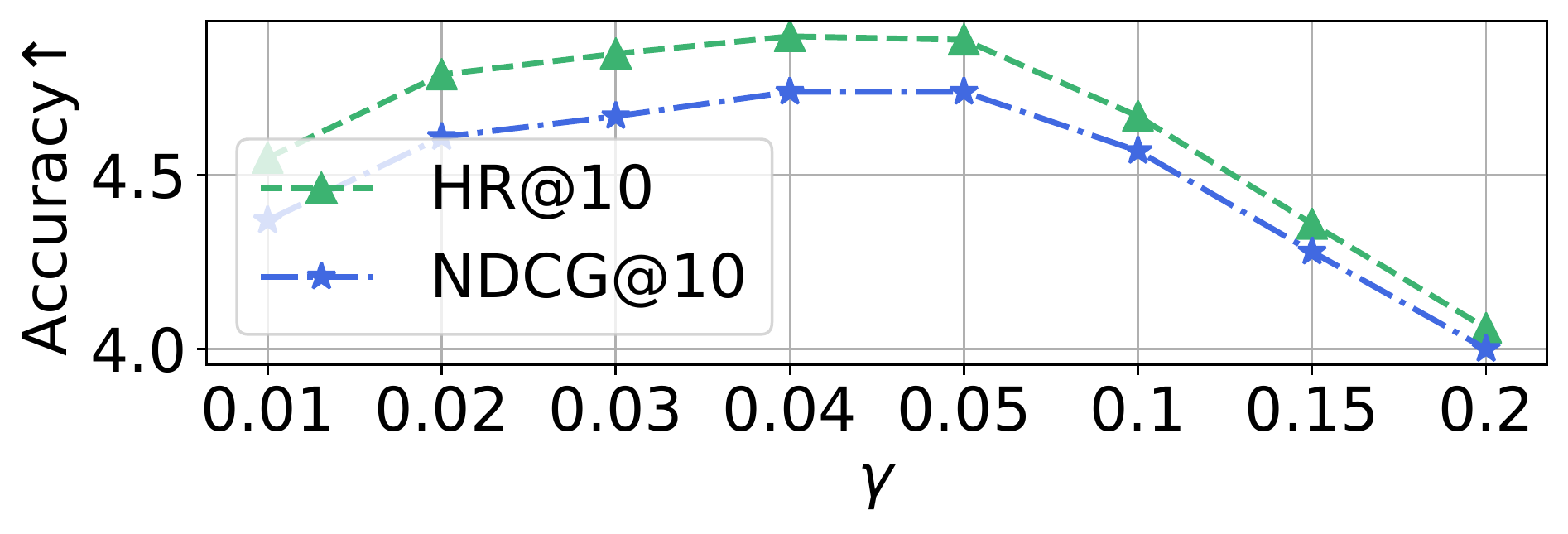}}
    \caption{Performance comparison w.r.t. the only hyperparameter \boldsymbol{$\gamma$} of SVD-AE. More results in other datasets are in the \emph{Appendix~\ref{app:gamma}}.}
    \label{fig:sensitivity}
\end{figure}

\begin{figure}[t]
    \centering
    \includegraphics[width=0.86\columnwidth]{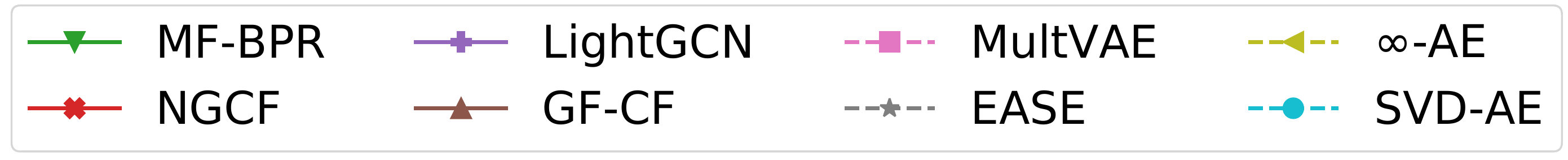}
    \subfigure[HR@10 (Gowalla)]{\includegraphics[width=0.43\columnwidth]{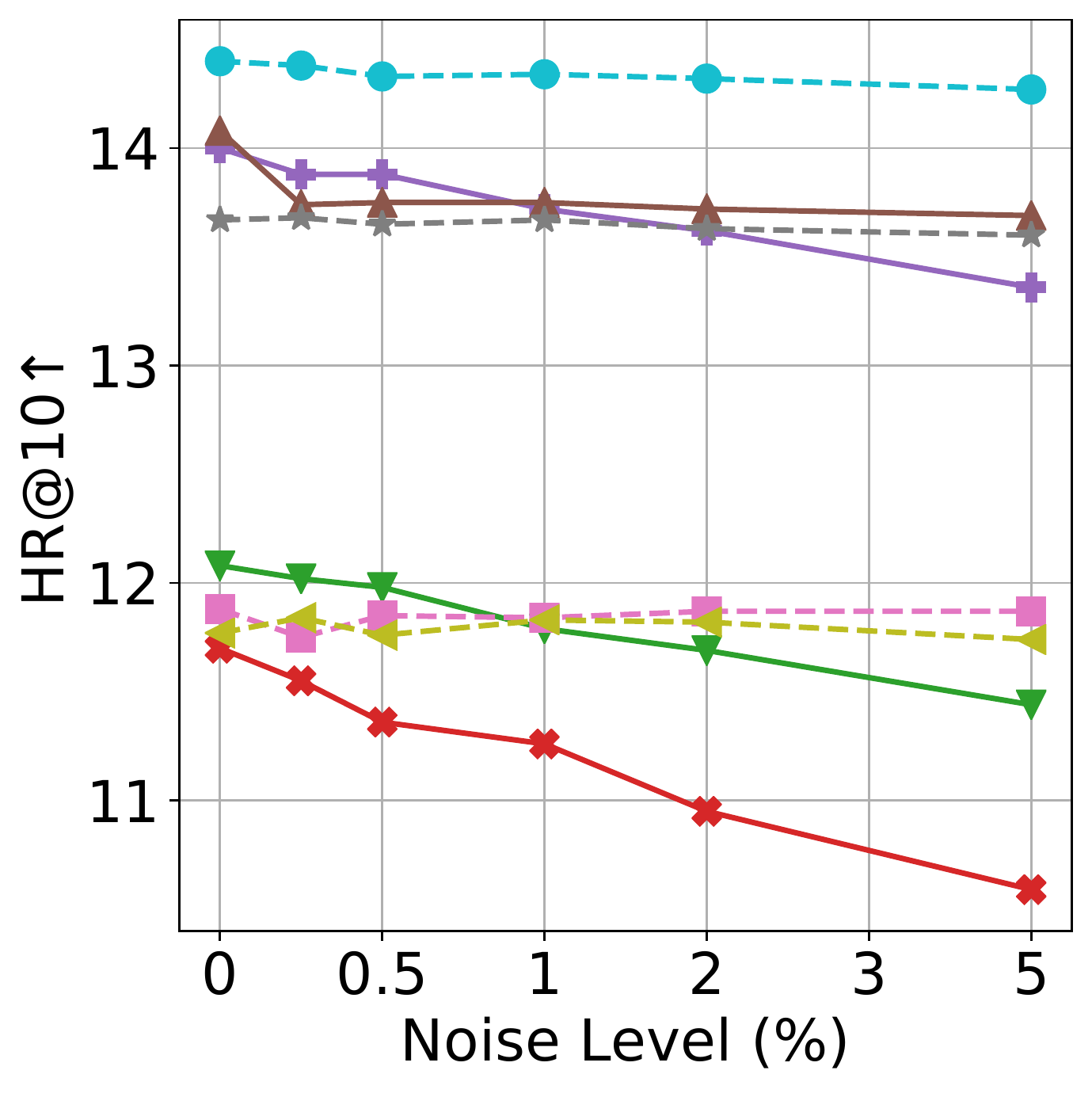}}
    \subfigure[NDCG@10 (Gowalla)]{\includegraphics[width=0.43\columnwidth]{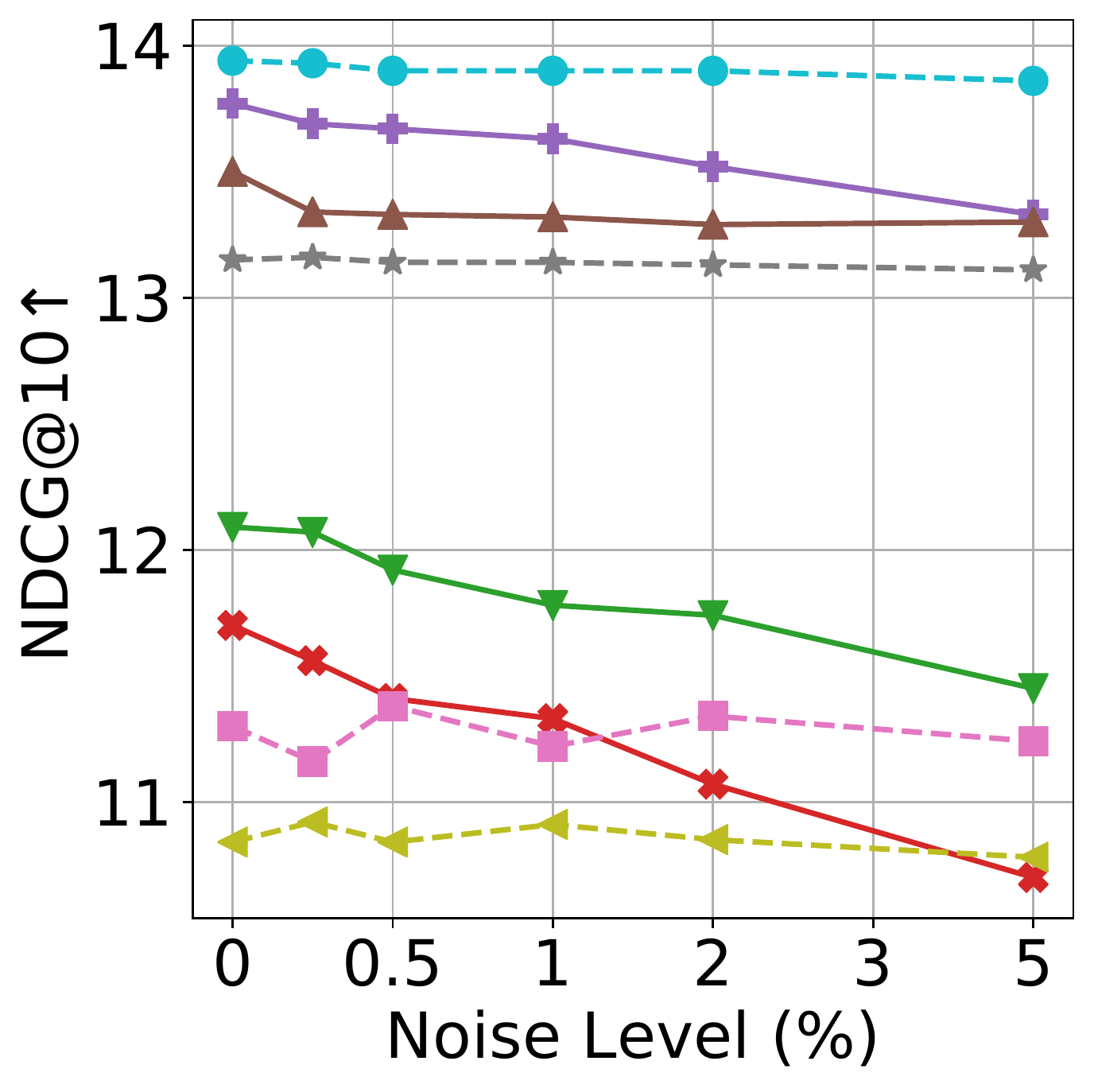}}
    \subfigure[HR@10 (Yelp2018)]{\includegraphics[width=0.43\columnwidth]{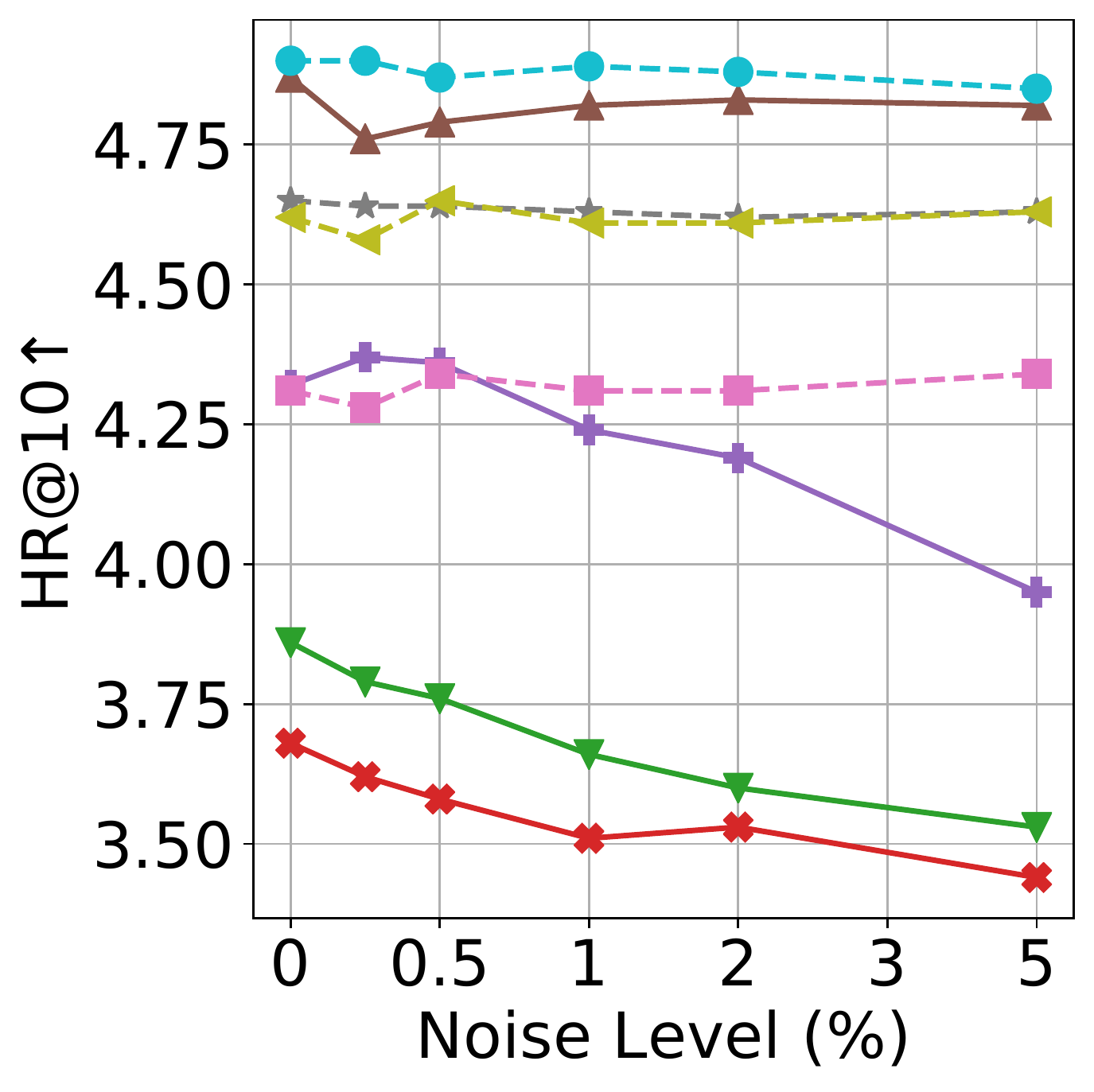}}
    \subfigure[NDCG@10 (Yelp2018)]{\includegraphics[width=0.43\columnwidth]{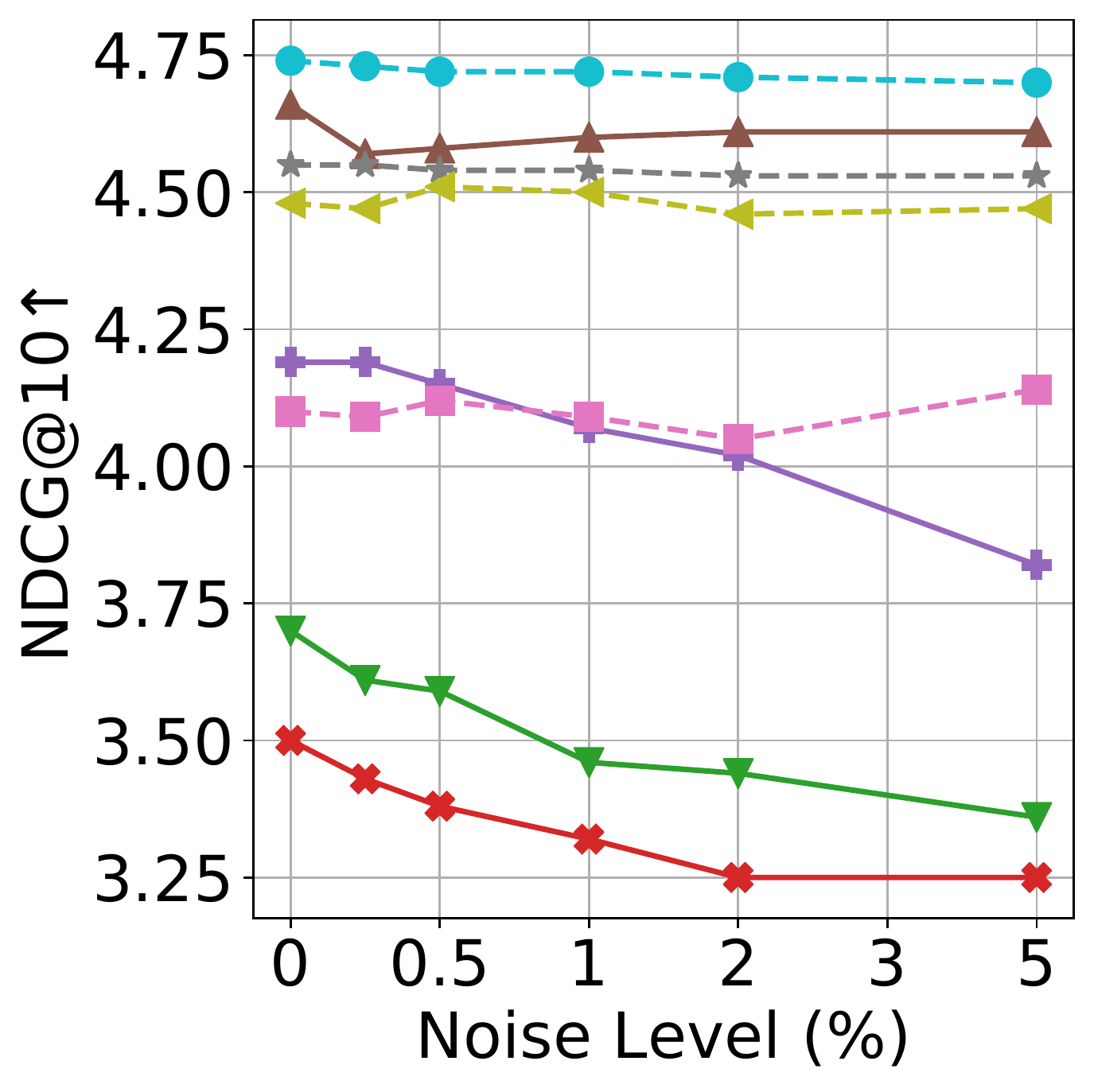}}
    \vspace{-0.3cm}
    \caption{Robustness evaluation against noise level. Solid line for non-AE models, dashed line for AE-based models.}
    \label{fig:robustness}
\end{figure}

\subsection{Robustness Analyses}
\paragraph{Robustness on Observed Noise.} Similar to Figure~\ref{fig:sensitivity}, here we investigate the influence of the rank parameter $\gamma$ of our SVD-AE. On all datasets, we examine the effectiveness of SVD-AE with different $\gamma$.
We observe that the MSE loss decreases as $\gamma$ (consequently, $m$) increases. This is because the truncated SVD is performed as closely as possible to the original matrix as $m$ increases. We can see, nevertheless, that recommendation performance is not always enhanced by using larger $m$. As the MSE loss decreases, the recommendation performance increases up to a certain level. However, beyond a certain threshold, we can observe a decline in the recommendation performance instead. This is because the noise from the rating matrix would be included in the reconstructed matrix if $m$ is too large. Such an occurrence further demonstrates that minimizing MSE loss does not ensure that recommendations may work better, which is consistent with what we discovered in the previous experiments on existing AE-based recommender systems which is illustrated in Figure~\ref{fig:overfitting_baselines}. 
Furthermore, it supports our low-rank inductive bias.

\paragraph{Robustness on Unobserved Noise.} In this part, we study the noise robustness of various CF methods. To evaluate the robustness of the proposed method and baselines, we show the performance of all methods on Gowalla and Yelp2018 with different noise levels in Figure~\ref{fig:robustness}. To simulate the noise, we generate random user-item interactions from 0.5\% to 5\% of the dataset size, thereby adjusting the amount of noise to different dataset sizes. 

We explore the robustness of four AE models, including MultVAE, EASE, $\infty$-AE, and our SVD-AE, and four non-AE models, including MF-BPR, NGCF, LightGCN, and GF-CF. 
Note that NeuMF is excluded for better visualization due to its significantly lower performance. In Figure~\ref{fig:robustness}, we can observe that our method can continue to perform steadily and competitively on noisy data of different degrees. We can see the greatest HR@10 and NDCG@10 are maintained regardless of the noise levels for Gowalla and Yelp2018.


On the other hand, a noteworthy feature is a significant disparity in noise robustness between AE and non-AE models. We discover that non-AE methods that disregard noise, such as MF-BPR, NGCF, and LightGCN, perform noticeably worse than the others when the noisy interactions are added. 
As a result, their HR@10 and NDCG@10 fall more significantly as the noise level increases. 
We can show that, among non-AE models, GF-CF is the most stable. 
We believe this is the indirect effect of using truncated SVD, although GF-CF does not use it to avoid the overfitting problem.

Additionally, we can observe that AE-based models, including MultVAE, EASE, and $\infty$-AE, are consistently stable for various noise levels.
However, as shown in Figure~\ref{fig:robustness}, they perform less reliably than SVD-AE. 
Notably, they also often perform worse than LightGCN and GF-CF in the experiment on Gowalla as shown in Figures~\ref{fig:robustness} (a) and (b). 
Based on these observations, our SVD-AE provides the best recommendation accuracy while maintaining robust performance even with the addition of noise.

\section{Conclusion}
We proposed a simplified but efficient linear autoencoder based on SVD and featuring a closed-form solution for collaborative filtering. Our SVD-AE outperforms various recommender systems including AE-based and non-AE models. Furthermore, compared to other methods, our method's runtime complexity is remarkably low. We discovered that our low-rank inductive bias via the truncated SVD makes it possible to obtain a closed-form solution and to effectively improve the robustness of the recommendation. Given the noisy nature of the rating matrix, we also discussed the overfitting issues with current recommender systems and presented the superior noise robustness of AE-based techniques, including SVD-AE. Our approach achieves the best balance between recommendation accuracy, computation time, and noise robustness. We expect much follow-up research work based on our observations of the new closed-form solution to the recommendation problem, in addition to the recent successes of developing lightweight methods with closed-form solutions.

\section*{Acknowledgements}
This work was partly supported by the Korea Advanced Institute of Science and Technology (KAIST) grant funded by the Korea government (MSIT) (No. G04240001, Physics-inspired Deep Learning) and Institute for Information \& Communications Technology Planning \& Evaluation (IITP) grants funded by the Korea government (MSIT) (No.2020-0-01361, Artificial Intelligence Graduate School Program (Yonsei University)).

\bibliographystyle{named}
\bibliography{ref}

\clearpage
\appendix

\section{Extensive and Detailed Related Work}\label{app:related_work}
\paragraph{Matrix Factorization.}
Collaborative filtering focuses on finding out the user preference from previous user-item interactions. A common paradigm for CF is to represent the user and item as learnable embeddings and optimize them based on historical interactions. The user $u$'s rating of item $i$ is estimated as the dot product of the user embedding $\mathbf e_{u}$ and the item embedding $\mathbf e_{i}$, (i.e., $\mathbf e_{u}^T\mathbf{e}_{i}$). 

Latent variable models are effective techniques for modeling user preferences and providing reliable recommendations~\cite{hofmann2004latent,ning2011slim,kabbur2013fism,agarwal2010flda,Hu08WRMF,Koren09MF,rendle2012bpr}. In general, these methods embed users and items into latent spaces and identify similarities between them. The most common type of latent variable model used in recommender systems is matrix factorization (MF)~\cite{agarwal2010flda,Hu08WRMF,Koren09MF,rendle2012bpr}. This approach uses MF to find user and item similarities and makes predictions based on these similarities. Traditional MF-based CF techniques merely use dot products to capture the behavior patterns in user-item interactions. Learning of the user behavior is initially suggested using SVD. It works by decomposing a matrix into constituent arrays of feature vectors corresponding to each user and each item, enabling the system to calculate similarities between users and items and to make predictions based on those similarities. These methods developed into other advanced MF methods~\cite{rao2015collaborative,yang2018hop,chen2020efficient} to effectively learn complex user and item representations. 
However, MF may not be able to fully capture the complicated and non-linear structure of interaction data by simply multiplying latent features linearly~\cite{he2017neural}. 
Accordingly, deep learning has received a lot of attention in recent literature because it has many advantages over conventional methods~\cite{he2017neural,liang2018variational}. Neural Collaborative Filtering (NCF)~\cite{he2017neural}, for instance, investigated neural network architectures for CF that benefit from extending MF to non-linear transformations.

\paragraph{Graph-based Collaborative Filtering.}
As the user-item interactions can be represented as the user-item bipartite graph or the item-item graph, graph convolutional networks (GCNs) have been used in recent studies~\cite{berg2017graph,ying2018graph,wang2019neural,wang2020disentangled} to learn high-order connectivity from user-item interactions. By directly capturing the collaborative signals in user-item interactions, these GCN-based methods can efficiently learn behavioral patterns between users and items. In general, they model a set of user-item interactions as a user-item graph and perform non-linear activation and feature transformation. GG-MC~\cite{berg2017graph} is a graph autoencoder framework based on passing differentiable messages on a bipartite interaction graph. PinSage~\cite{ying2018graph} is a data-efficient GCN that combines efficient random walks and GCNs to produce node embeddings that contain both the graph's structure and its node features. NGCF~\cite{wang2019neural} propagates user and item embeddings by layer combination to capture the collaborative signals among users and items. DGCF~\cite{wang2020disentangled} generates disentangled representations by viewing user-item relationships at the finer granularity of user intents.

However, recent works~\cite{wu2019simplifying,he2020lightgcn,chen2020revisiting,shen2021powerful,peng2022svd} show that the performance of GCN-based CF methods can be improved by removing the non-linear activation functions and feature transformation. They argue that the efficacy of GCNs is due to neighborhood aggregation. By eliminating non-linear activation and feature transformation, LightGCN~\cite{he2020lightgcn} outperformed NGCF in both accuracy and efficiency. Especially, GF-CF~\cite{shen2021powerful} viewed CF from the perspective of graph signal processing and it simply requires a SVD process and adjacency matrix multiplication to estimate rating scores. A linear graph filter and an ideal low-pass graph filter were combined to form the GF-CF. SVD-GCN~\cite{peng2022svd} also offered a simplified GCN learning paradigm by using the truncated SVD to replace the core design of GCN-based approaches.

\paragraph{Autoencoder-based Collaborative Filtering.}
An autoencoder~\cite{bengio2006greedy} is typically implemented as a neural network that learns a hidden representation of an input and then reconstructs it. Common approaches to CF are based on the idea of autoencoding features from the preference matrix. We can examine scores for all potential items of interest which is unseen by users by decoding preference histories. AutoRec~\cite{sedhain2015autorec} is a compact and efficiently trainable autoencoder framework for CF. CDAE~\cite{wu2016collaborative} used a standard denoising autoencoder for recommendations. Hybrid CF methods, which combine deep learning techniques and latent variable modeling, have also been intensively investigated as a means of dealing with the user-item matrix's growing sparsity. CDL~\cite{wang2015collaborative} presented a Bayesian formulation of a stacked denoising autoencoder~\cite{vincent2010stacked} for combined deep representation learning of content and CF. AutoSVD++~\cite{zhang2017autosvd++} generalized contractive autoencoder to matrix factorization.

The introduction of the variational autoencoder (VAE) framework has led to the development of advanced AE-based CF methods. A Bayesian generative model called CVAE~\cite{li2017collaborative} utilized the item latent representations from VAE along with the ratings. In contrast, MultVAE~\cite{liang2018variational} extended variational autoencoders to model user-item feedback data effectively using multinomial likelihood. SVAE~\cite{sachdeva2019sequential} applied variational autoencoders to the case of sequential recommendation and presented a recurrent variant of the VAE. They captured the temporal dependencies between the consumption sequences by using the recurrent encoder. To build a more individualized CF model, MacridVAE~\cite{ma2019learning} suggested learning the disentangled representation of user behavior. RecVAE~\cite{shenbin2020recvae} enhanced MVAE by providing a composite prior distribution for the latent code. BiVAE~\cite{truong2021bilateral} constructed an autoencoder that collectively encodes both users and items. MD-CVAE~\cite{zhu2022mutually} incorporated latent item embeddings to the user-oriented autoencoder.

To build lightweight recommender systems that infer user preferences, several recent autoencoder approaches used linear regression~\cite{steck2019embarrassingly,sachdeva2022infinite}. EASE~\cite{steck2019embarrassingly} suggested a shallow autoencoder that simply calculates an item-item similarity matrix using ordinary least squares regression with a closed form solution. They also used Lagrangian multipliers to constrain the self-similarity of each item in the input and output layer to zero. Recently, $\infty$-AE~\cite{sachdeva2022infinite} proposed training infinitely-wide neural networks to create an autoencoder with infinitely-wide bottleneck layers. The NTK, which they used, offers a simple recommendation model with a single hyper-parameter and a closed-form solution. RLAE and RDLAE~\cite{moon2023s} also employed linear autoencoders. They tuned for L2 regularization with random dropout, which changed the degree of diagonal constraints.

\section{Details on Section~\ref{sec:svd-ae}}\label{app:proof}
\subsection{Proof of Theorem 1}
Let $\tilde{\mathbf{R}} \in \mathbb{R}^{m \times n}$ and its pseudo-inverse is $\tilde{\mathbf{R}}^+ = (\tilde{\mathbf{R}}^T\tilde{\mathbf{R}})^{-1}\tilde{\mathbf{R}}^T$. Suppose its SVD is $\tilde{\mathbf{R}} = \mathbf{Q}\mathbf{\Sigma}\mathbf{V}^T$ where $\mathbf{Q}, \mathbf{V}$ are both orthogonal matrices and $\mathbf{\Sigma}$ is a diagonal matrix containing the singular values. Then the pseudo-inverse of $\tilde{\mathbf{R}}$ is $\tilde{\mathbf{R}}^+ = \mathbf{V}\mathbf{\Sigma}^+\mathbf{Q}^T$. 
\begin{proof}
We can derive $\tilde{\mathbf{R}}^+ = \mathbf{V}\mathbf{\Sigma}^+\mathbf{Q}^T$ as follows:
\begin{align}
\tilde{\mathbf{R}}^+ &= (\tilde{\mathbf{R}}^T\tilde{\mathbf{R}})^{-1}\tilde{\mathbf{R}}^T \\
 &= (\mathbf{V}\mathbf{\Sigma}^T\mathbf{Q}^T\mathbf{Q}\mathbf{\Sigma}\mathbf{V}^T)^{-1}\mathbf{V}\mathbf{\Sigma}\mathbf{Q}^T \\
 &= (\mathbf{V}\mathbf{\Sigma}^T\mathbf{\Sigma}\mathbf{V}^T)^{-1}\mathbf{V}\mathbf{\Sigma}\mathbf{Q}^T \\
 &= (\mathbf{V}^T)^{-1}(\mathbf{\Sigma}^2)^{-1}\mathbf{V}^{-1}\mathbf{V}\mathbf{\Sigma}^T\mathbf{Q}^T \\
 &= \mathbf{V}\mathbf{\Sigma}^{-2}\mathbf{\Sigma}\mathbf{Q}^T\\
 &= \mathbf{V}\mathbf{\Sigma}^{-1}\mathbf{Q}^T \\
 &= \mathbf{V}\mathbf{\Sigma}^{+}\mathbf{Q}^T
\end{align}

since $\mathbf{\Sigma}^T = \mathbf{\Sigma}$ and $\mathbf{\Sigma}^+ = (\mathbf{\Sigma}^T\mathbf{\Sigma})^{-1}\mathbf{\Sigma}^T = \mathbf{\Sigma}^{-1}$.
\end{proof}

\begin{proof}
We show $\tilde{\mathbf{R}}^+ = \mathbf{V}\mathbf{\Sigma}^+\mathbf{Q}^T$ satisfies the four properties, known as the Moore–Penrose conditions:
\begin{align}
\tilde{\mathbf{R}}\tilde{\mathbf{R}}^+\tilde{\mathbf{R}} &= \mathbf{Q}\mathbf{\Sigma}\mathbf{V}^T \cdot \mathbf{V}\mathbf{\Sigma}^+\mathbf{Q}^T \cdot \mathbf{Q}\mathbf{\Sigma}\mathbf{V}^T = \mathbf{Q}\mathbf{\Sigma}\mathbf{V}^T = \tilde{\mathbf{R}} \\
\tilde{\mathbf{R}}^+\tilde{\mathbf{R}}\tilde{\mathbf{R}}^+ &= \mathbf{V}\mathbf{\Sigma}^+\mathbf{Q}^T \cdot \mathbf{Q}\mathbf{\Sigma}\mathbf{V}^T \cdot \mathbf{V}\mathbf{\Sigma}^+\mathbf{Q}^T = \mathbf{V}\mathbf{\Sigma}^+\mathbf{Q}^T = \tilde{\mathbf{R}}^+ \\
\tilde{\mathbf{R}}\tilde{\mathbf{R}}^+ &= \mathbf{Q}\mathbf{\Sigma}\mathbf{V}^T \cdot \mathbf{V}\mathbf{\Sigma}^+\mathbf{Q}^T = \mathbf{Q}\mathbf{\Sigma}\mathbf{\Sigma}^+\mathbf{Q}^T \quad \text{(symmetric)}\\
\tilde{\mathbf{R}}^+\tilde{\mathbf{R}} &= \mathbf{V}\mathbf{\Sigma}^+\mathbf{Q}^T \cdot \mathbf{Q}\mathbf{\Sigma}\mathbf{V}^T = \mathbf{V}\mathbf{\Sigma}^+\mathbf{\Sigma}\mathbf{V}^T \quad \text{(symmetric)}.
\end{align}
\end{proof}

\subsection{Proof of Theorem 2}
\begin{proof}
Suppose $\mathbf{B} = \tilde{\mathbf{R}}^+\mathbf{R}$ is a minimum norm solution to $\tilde{\mathbf{R}}\mathbf{B} = \mathbf{R}$. To show that it has the minimum norm, for any solution $\mathbf{B}$, consider its orthogonal decomposition via $\tilde{\mathbf{R}}^+\tilde{\mathbf{R}} \in \mathbb{R}^{n \times n}$:  

\begin{equation}
    \mathbf{B} = (\tilde{\mathbf{R}}^+\tilde{\mathbf{R}})\mathbf{B} + (\mathbf{I} - \tilde{\mathbf{R}}^+\tilde{\mathbf{R}})\mathbf{B} = \tilde{\mathbf{R}}^+\mathbf{R} + (\mathbf{I} - \tilde{\mathbf{R}}^+\tilde{\mathbf{R}})\mathbf{B}.
\end{equation}

It follows that

\begin{equation}
    \|\mathbf{B}\|^2 = \|\tilde{\mathbf{R}}^+\mathbf{R}\|^2 + \|(\mathbf{I} - \tilde{\mathbf{R}}^+\tilde{\mathbf{R}})\mathbf{B}\|^2 \geq \|\tilde{\mathbf{R}}^+\mathbf{R}\|^2.
\end{equation}

This shows that $\|\mathbf{B}\| \geq \|\tilde{\mathbf{R}}^+\mathbf{R}\|$.

\end{proof}

\subsection{Proof of Theorem 3}

Let $\tilde{\mathbf{R}} \in \mathbb{R}^{m \times n}$ given by $\tilde{\mathbf{R}}=\mathbf{D}_U^{-\frac{1}{2}}\mathbf{R}\mathbf{D}_I^{-\frac{1}{2}}$. Consider the matrix $\tilde{\mathbf{A}} = \tilde{\mathbf{R}}^T\tilde{\mathbf{R}}$. It is a symmetric $n \times n$ matrix, so its eigenvalues are real. We can prove $\sigma \geq 0$ where $\sigma$ is an singular value of $\tilde{\mathbf{R}}$ by showing $\lambda \geq 0$ where $\lambda$ is an eigenvalue of $\tilde{\mathbf{A}}$: 
\begin{proof}
Let $x$ be an eigenvector of $\tilde{\mathbf{A}}$ with eigenvalue $\lambda$. We compute that
\begin{equation}
    \|\tilde{\mathbf{R}}x\|^2 = (\tilde{\mathbf{R}}x)^T(\tilde{\mathbf{R}}x) = x^T\tilde{\mathbf{A}}x = x^T(\lambda x) = \lambda x^Tx = \lambda \|x\|^2.
\end{equation}

Since $\|\tilde{\mathbf{R}}x\|^2 \geq 0$, we can prove $\lambda \|x\|^2 \geq 0$ from the above equation. Then we can deduce that $\tilde{\mathbf{A}}$ is positive semi-definite and $\lambda \geq 0$. Let $\{\lambda_1, \lambda_2, ..., \lambda_M\}$ indicate the eigenvalues of $\tilde{\mathbf{A}}$. If we order them to $\lambda_1 \geq ... \geq \lambda_M \geq 0$ and let $\sigma_i = \sqrt{\lambda_i}$, so that $\sigma_1 \geq ... \geq \sigma_M \geq 0$.
\end{proof}

We now show $\sigma \leq 1$:
\begin{proof}
For any $x$ such that $\|x\|_2 = 1$, we have 
\begin{equation}
    1 - x^T\tilde{\mathbf{A}}x = x^T(\mathbf{I} - \tilde{\mathbf{A}})x = \sum_{(u,i)}(\frac{x_u}{\sqrt{d_u}} - \frac{x_i}{\sqrt{d_i}})^2 \geq 0,
\end{equation} where $u \in U, i \in I$. The above equation shows that $-1 \leq x^T(\mathbf{I}-\tilde{\mathbf{A}})x \leq 1$. Using the vector $x = \mathbf{D}^\frac{1}{2}\mathbf{1}$, we show

\begin{equation}
\tilde{\mathbf{A}}x = \mathbf{D}^{-\frac{1}{2}}\mathbf{A}\mathbf{D}^{-\frac{1}{2}}\mathbf{D}^\frac{1}{2}\mathbf{1} = \mathbf{D}^{-\frac{1}{2}}\mathbf{A}\mathbf{1} = \mathbf{D}^{-\frac{1}{2}}\text{diag}(\mathbf{D}) = \mathbf{D}^\frac{1}{2}\mathbf{1}.
\end{equation}

This implies that $\lambda_{max}$, the largest eigenvalue of $\tilde{\mathbf{A}}$ is 1. Thus, the largest singular value of $\tilde{\mathbf{R}}$ is 1 since $\sigma_i = \sqrt{\lambda_i}$.

\end{proof}

\section{Detailed Experimental Settings}

\subsection{Environmental Setup.}
Our software and hardware setup is as follows: \textsc{Python} 3.9.15, \textsc{PyTorch} 1.11.0, \textsc{CUDA} 11.4, and \textsc{NVIDIA} Driver 470.129.06, and i9 CPU, and \textsc{NVIDIA Quadro RTX 8000}. 

\subsection{Detail of Datasets}\label{app:dataset}
Table~\ref{tbl:stat} shows the detailed statistics of datasets. In our experiments, we utilize 4 benchmark datasets~\cite{cho2011friendship,harper2015movielens}: Gowalla, Yelp2018, ML-1M, and ML-10M.
Following ~\cite{wang2019neural,he2020lightgcn,shen2021powerful,sachdeva2022infinite}, all user-item interactions are treated as positive, so that all the datasets are implicit feedback. 
For preprocessing, Gowalla and Yelp2018 were preprocessed with a 10-core setting~\cite{wang2019neural}, while ML-1M and ML-10M were published with a 5-core setting.
It should be noted that we use the same datasets as in the baselines for a fair comparison.

\begin{table}[h]
    \centering
    \begin{tabular}{l rrrr}
    \toprule
    Dataset              & \#User & \#Item & \#Interaction & Density \\ \midrule
    Gowalla              & 29,858 & 40,981 & 1,027,370     & 0.0008   \\
    Yelp2018             & 31,668 & 38,048 & 1,561,406     & 0.0013   \\
    ML-1M                & 6,040  & 3,706  & 1,000,209     & 0.0447   \\
    ML-10M               & 69,878 & 10,677 & 10,000,054    & 0.0134   \\ \bottomrule
    \end{tabular}
    \caption{Statistics of experimented datasets}
    \label{tbl:stat}
\end{table}

\subsection{Detail of Baselines}\label{app:baseline}
A brief description of the baseline models we used is as follows:
\begin{enumerate}
    \item MF-BPR~\cite{rendle2012bpr} is a factorization method with the BPR loss.
    \item NeuMF~\cite{he2017neural} is a deep learning model that learns user and item embeddings.
    \item NGCF~\cite{wang2019neural} is a GCN model with non-linear activations.
    \item LightGCN~\cite{he2020lightgcn} is a lightweight GCN-based model only with a linear neighbourhood aggregation.
    \item GF-CF~\cite{shen2021powerful} is a computationally efficient CF method that uses graph filters.
    \item MultVAE~\cite{liang2018variational} redesigns LightGCN on top of neural ordinary differential equations.
    \item EASE~\cite{steck2019embarrassingly} is a linear model using an auto-encoder and least squares regression.
    \item $\infty$-AE~\cite{sachdeva2022infinite} is an infinitely-wide autoencoder which utilizes the NTK.
\end{enumerate}

\subsection{Hyperparameter Configurations}\label{app:hyperparam}
As shown in Table~\ref{tbl:range}, we execute baselines based on their official codes and
suggested hyperparameter ranges. For our SVD-AE, we search for $\gamma$ in the range of $[0.01, 0.02, 0.03, 0.04, 0.05 ]$ with the validation set with HR@10. Note that $\gamma$ is the only hyperparameter of our method and $\gamma = 0.04$ is the optimal value for all datasets.

\begin{table}[h]
\centering
\begin{tabular}{lll}
\toprule
Hyperparameter                 & Model       & Search Range                       \\ \midrule
\multirow{4}{*}{Latent size}   & MF-BPR      & \multirow{4}{*}{\{32, 64, 128\}}   \\
                               & NeuMF       &                                    \\
                               & NGCF        &                                    \\
                               & LightGCN    &                                    \\ \midrule
\multirow{5}{*}{Learning rate} & MF-BPR      & \multirow{5}{*}{\{1e-3, 1e-4\}}    \\
                               & NeuMF       &                                    \\
                               & NGCF        &                                    \\
                               & LightGCN    &                                    \\
                               & MultVAE     &                                    \\ \midrule
\multirow{4}{*}{Dropout rate}  & MF-BPR      & \multirow{4}{*}{\{0.1, 0.3, 0.5\}} \\
                               & NeuMF       &                                    \\
                               & NGCF        &                                    \\
                               & MultVAE     &                                    \\ \midrule
\multirow{2}{*}{$\lambda$}     & EASE        & \{1, 10, 100, 1000, 10000\}        \\
                               & $\infty$-AE & \{0, 1, 5, 20, 50, 100\}           \\ \midrule
$\alpha$                       & GF-CF       & \{0.1, 0.3, 0.5, 0.7, 0.9\}        \\ \bottomrule
\end{tabular}
\caption{List of hyperparameters grid-searched for baselines}
\label{tbl:range}
\end{table}

\pagebreak
\section{More Visualization in Robustness Analyses}\label{app:gamma}

\begin{figure}[h]
    \centering
    \subfigure[MSE (ML-1M)]{\includegraphics[width=0.48\columnwidth]{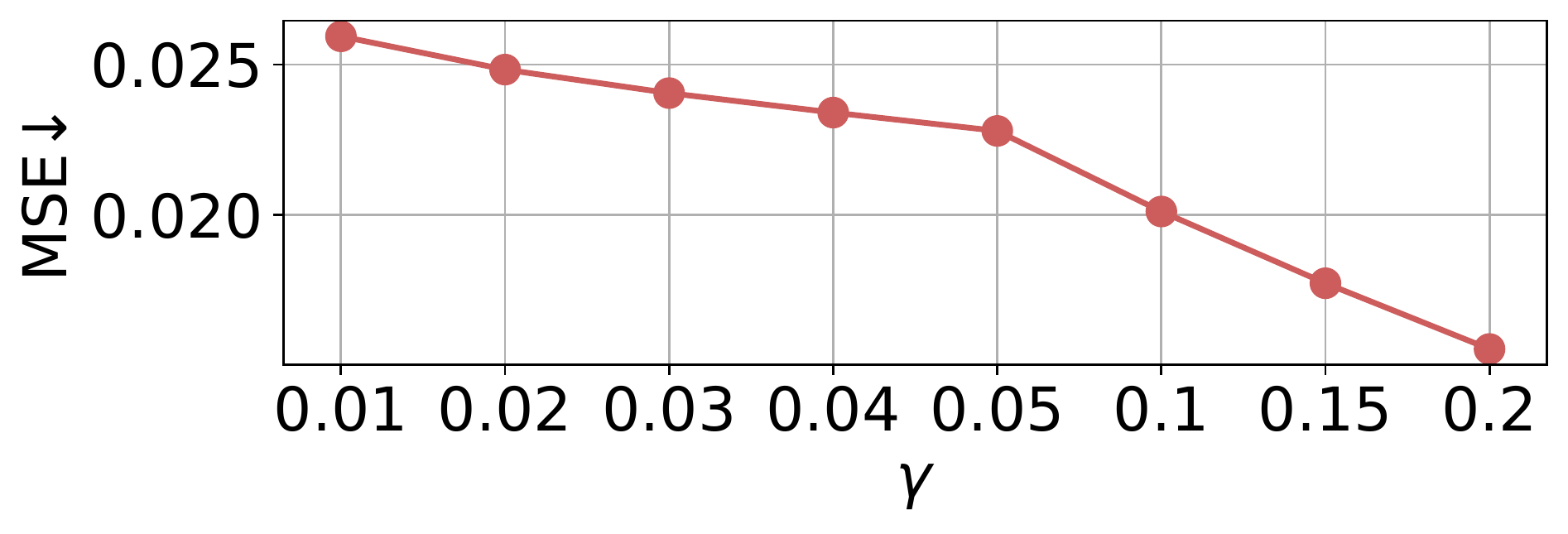}}
    \subfigure[Accuracy (ML-1M)]{\includegraphics[width=0.48\columnwidth]{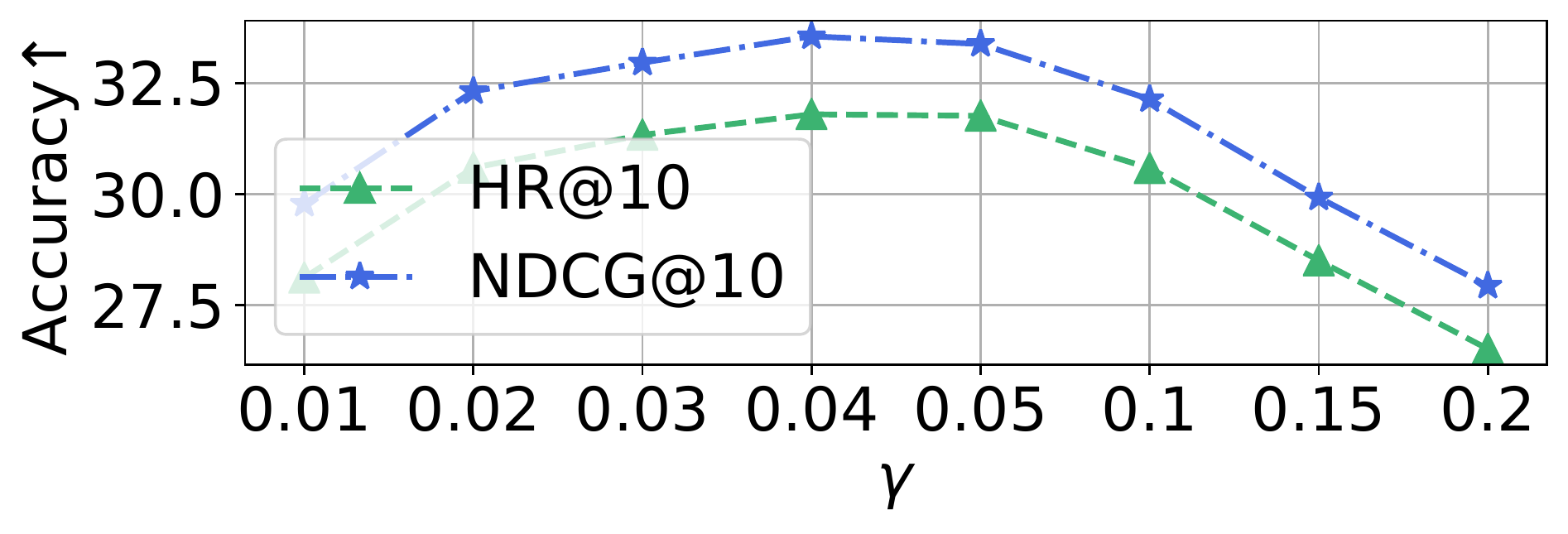}}
    \subfigure[MSE (ML-10M)]{\includegraphics[width=0.48\columnwidth]{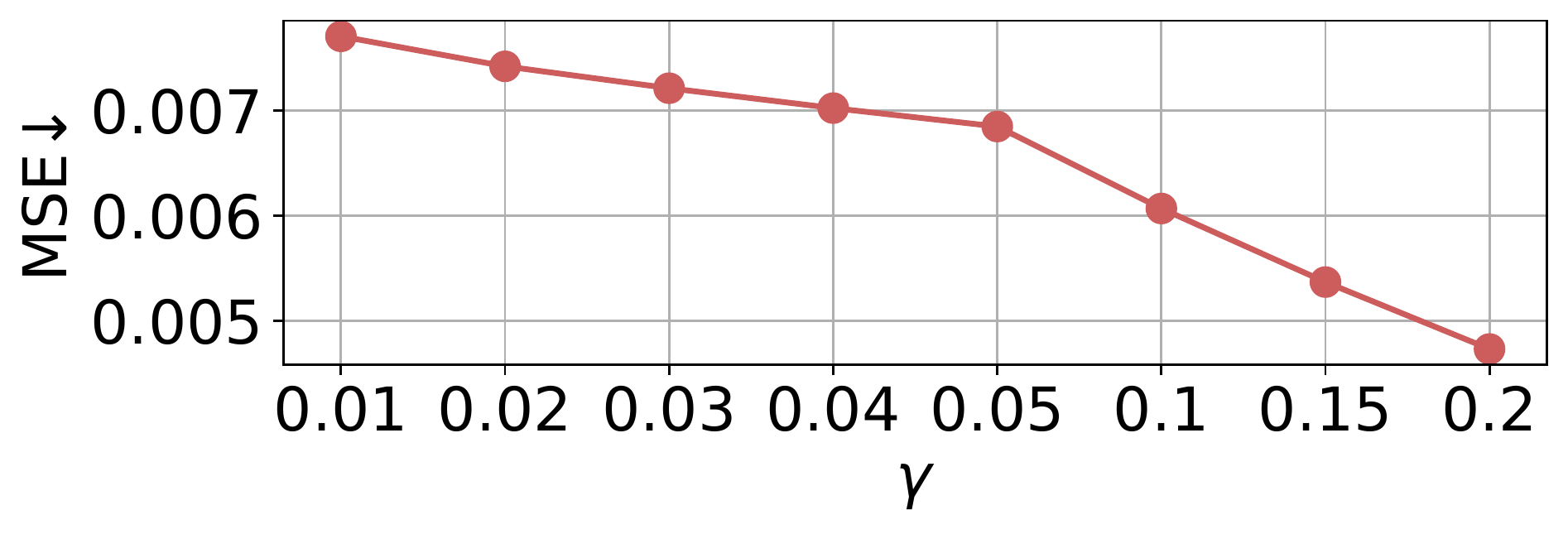}}
    \subfigure[Accuracy (ML-10M)]{\includegraphics[width=0.48\columnwidth]{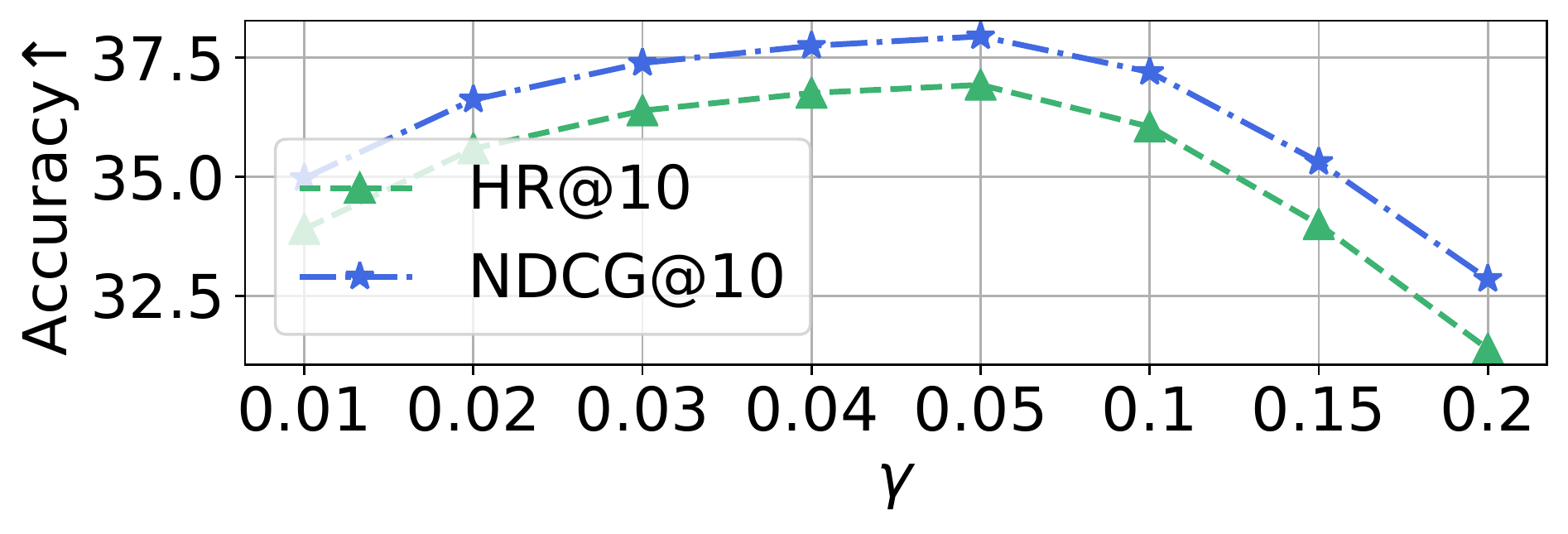}}
    \vspace{-0.3cm}
    \caption{Performance comparison w.r.t. the only hyperparameter \boldsymbol{$\gamma$} of SVD-AE. Note that \boldsymbol{$\gamma$} is used to obtain the rank parameter \boldsymbol{$m = \lfloor \gamma \times \text{min}(|U|, |I|)\rceil$} for the truncated SVD, where \boldsymbol{$\lfloor ... \rceil$} function denotes the rounding function. Higher rank guarantees the lowest MSE loss but it does not always yield the best recommendation performance, which justifies the presence of noise and low-rank inductive bias. }
    \label{fig:sensitivity-app}
\end{figure}

\clearpage
\end{document}